\documentclass{aa}  

\usepackage{graphicx}
\usepackage{txfonts}
\usepackage[dvipsnames]{xcolor}
\usepackage{gensymb}
\usepackage{hyperref}
\hypersetup{
    colorlinks  =  true,
    citecolor   =  blue,
}

\begin{document}

   \title{Confirming NGC 6231 as the parent cluster of the runaway high-mass X-ray binary HD 153919/4U 1700-37 with \textit{Gaia} DR2\thanks{The parallaxes and proper motions used in this paper have been updated to EDR3.}}

   \subtitle{}

   \author{Vincent van der Meij
          \inst{1}
          \and
          Difeng Guo
          \inst{1}
          \and
          Lex Kaper
          \inst{1}
          \and
          Mathieu Renzo
          \inst{1,2,3}
          }

   \institute{Anton Pannekoek Institute for Astronomy, University of Amsterdam,
              Science Park 904, 1098 XH, Amsterdam, The Netherlands\\
              \email{L.Kaper@uva.nl}
         \and 
         	Center for Computational Astrophysics, Flatiron Institute, New York, NY 10010, USA
         \and
         	Department of Physics, Columbia University, New York, NY 10027, USA\\
             }

   \date{Received September 00, 0000; accepted March 00, 0000}

\authorrunning{Van der Meij et al.} 
\titlerunning{NGC6231 the parent cluster of HMXB 4U1700-37}

 
  \abstract
   {A significant fraction (10-20\%) of the most massive stars move through space with a high ($v \gtrsim 30$~km~s$^{-1}$) velocity. One of the possible physical explanations is that a supernova in a compact binary system results in a high recoil velocity of the system. If the system remains bound, it can be subsequently observed as a spectroscopic binary (SB1), a high-mass X-ray binary, a compact binary, and finally a gravitational-wave event.}
   {If such a system is traced back to its parent cluster, binary evolution models can be tested in great detail.}
   {The \textit{Gaia} proper motions and parallaxes are used to demonstrate that the high-mass X-ray binary HD~153919/4U~1700-37 originates from NGC~6231, the nucleus of the OB association Sco OB1.}
   {The O supergiant and its compact companion, of which the physical nature (a neutron star or a black hole) is unknown, move with a space velocity of $63 \pm 5$~km~s$^{-1}$ with respect to NGC 6231. The kinematical age of the system is $2.2 \pm 0.1$~Myr. The parallaxes and accurate proper motions in \textit{Gaia} DR2 were used to perform a membership analysis of NGC 6231; 273 members are identified, of which 268 have good quality photometry. The distance to NGC~6231 is $1.63 \pm 0.15$~kpc. Isochrone fitting results in an age of $4.7 \pm 0.4$~Myr and an extinction $A_V$ to the cluster of $1.7 \pm 0.1$. With the identification of NGC~6231 as the parent cluster, the upper limit on the age of the progenitor of 4U1700--37 at the moment of the supernova explosion is $3.0 \pm 0.5$~Myr.}
   {With these constraints, the evolutionary history of the system can be reconstructed with an initial mass of the progenitor of the compact object $>$ 60 \(\textup{M}_\odot\). The high mass, the extreme mass ratio, and short orbital separation of the system make it difficult to produce possible progenitor systems through population synthesis. We propose that the system experienced a Case A mass transfer phase before the supernova, which typically widens a binary. In order to create a progenitor system that does not merge, a lot of angular momentum must be lost from the system during the phase of mass transfer and/or an asymmetry in the supernova explosion provides a kick resulting in the observed orbital parameters. Given its current high space velocity and the derived evolutionary history, the compact object in the system is more likely to have received a large natal kick, which suggests that it is more likely a neutron star than a black hole. HD153919/4U1700--37 might be a prototype in the Milky Way for the progenitor of gravitational wave events such as GW190412.}
 

   \maketitle

\section{Introduction}
\label{chap:HMXB|sec:intro}
Massive stars (M~$>8-10$~M$_{\odot}$) are hot and luminous and evolve rapidly (lifetimes up to a few tens of mega year). At the end of their lives, massive stars collapse producing a supernova and/or a gamma-ray burst. The end product is a compact object: a neutron star (NS) or a black hole (BH). During the supernova, various (heavy) chemical elements are produced; the outflow enriches the interstellar medium providing the building blocks for future generations of stars. For a review on the pre-supernova evolution of massive single and binary stars, see \citet{Langer2012}. For a review on the explosion mechanisms of core-collapse supernovae, we refer readers to \citet{Janka2012}.

The majority of massive stars are in binary (or multiple) systems \citep{Sana2012}. In a close binary, a phase of mass transfer before the supernova explosion of the primary results in an inversion of the mass ratio with the secondary becoming the most massive star in the system. This provides the condition that the binary system may remain bound after the supernova explosion; the latter also depends on the details of the natal kick \citep[see, e.g.,][]{Kalogera1998}. From that moment on, the binary consists of an OB-type star and a compact object that subsequently evolves into a high-mass X-ray binary (HMXB). X-rays are produced by the accretion of material from either the OB-star wind or through Roche-lobe overflow onto the compact object. HMXBs represent an important phase in the evolution of massive (close) binaries toward the formation of gravitational-wave sources: the merging event of a NS+NS, NS+BH or a BH+BH system \citep{Belczynski2002}. For a review on massive binary evolution, see \citet{Tauris2006}.

High-mass X-ray binaries are unique laboratories to test accretion physics and stellar evolution. HMXBs also provide the opportunity to accurately determine the physical properties of both the massive star and the compact object. If the compact object is an X-ray pulsar, its orbital parameters can be measured with high precision such that, for example, the masses of both companions are obtained given the radial-velocity amplitude of the OB star and the system inclination, such as in eclipsing systems. The X-ray eclipse duration provides a measure of the radius of the OB star. With the ({\it Gaia}) parallax, the OB-star spectrum and an estimate of the interstellar extinction, the effective temperature and luminosity of the OB star are obtained. As a consequence, the stellar parameters of the OB stars in HMXBs are amongst the most accurately known \citep[cf.][]{Kaper2001}.

Typically, due to their short lifetime, massive stars go into supernova close to the region where they were born. 
However, some massive stars move with high (supersonic) velocity through the interstellar medium, the so-called OB runaway stars (and walkaway stars) \citep{Blaauw1961, Renzo2019}. They travel through interstellar space and when they explode as a supernova, they may have reached more remote gas-rich regions in our Galaxy where star formation can be triggered. Alternatively, they explode at a relatively high distance above the Galactic plane, such that the nuclearly enriched material can escape from the Galaxy.

\begin{figure*}[ht]
	\centering
	\includegraphics[width=\linewidth]{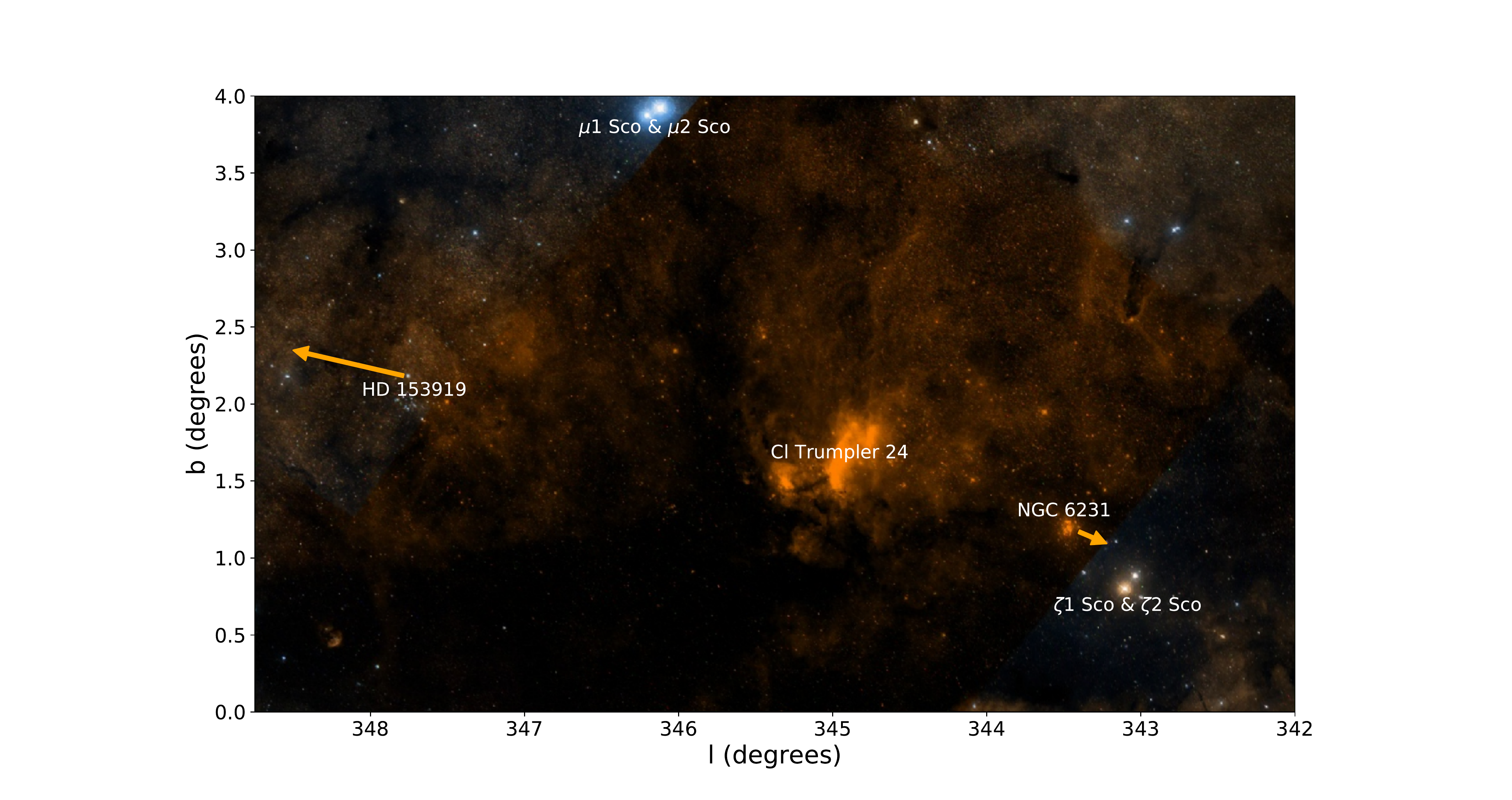}
	\caption{NGC 6231, the core ionizing cluster of the OB association Sco~OB1, and the high-mass X-ray binary HD 153919/ 4U1700-37. The orange arrows show the distance the objects will travel in 0.5~Myr. The nearby open cluster Cl Trumpler 24, surrounded by the Prawn Nebula (IC~4621), and 4 bright stars in the Scorpius constellation are labeled. Source: Aladin.}
	\label{fig:ScoOB1}
\end{figure*}

In the vast majority of cases, the progenitor of the compact object was the originally most massive star in the binary \citep{Pols1994}. This star depletes its hydrogen core faster than its companion, and expands into the supergiant phase first. If during this process the radius of the primary exceeds the Roche-lobe radius, it will transfer mass from its envelope onto the secondary. The secondary accretes hydrogen-rich material such that it becomes the most massive star in the system and the mixing of hydrogen into the core rejuvenates the star \citep{VanBever1998, Schneider2016, Hellings1983}. The primary becomes a helium star and produces a compact object after core collapse. If less than half of the total mass of the system is lost, the system can remain bound \citep{Boersma1961}, depending on the natal kick, and becomes a HMXB when the accretion rate onto the compact object results in an observable X-ray flux.

Most HMXBs are runaway systems \citep{VanDenHeuvel2000}; when they move supersonically through the interstellar medium, a wind-bow shock can be produced \citep{Kaper1997, Gvaramadze2011, Meyer2014, Prisegen2019}. \citet{Blaauw1961} proposed that an OB runaway obtains its high space velocity as a consequence of a supernova explosion in a binary. If the system remains bound, the binary will obtain a recoil velocity comparable to the orbital velocity of the supernova progenitor. In this scenario, HMXBs are predicted to be runaway systems (unless no or little mass is lost from the system).
However, the asymmetry of the supernova explosion likely introduces an additional velocity component (kick velocity) that may disrupt the binary, cf. \citet{Renzo2019}. An alternative scenario to produce OB runaways is through the dynamical ejection from a stellar cluster \citep{Poveda1967}. This is more likely to occur in a young cluster when the stellar density is still high. \cite{Hoogerwerf2000} demonstrated that both mechanisms are at work; see also \citet{Jones2020} and \citet{Jilinski2010}.

The time that passed since the OB-runaway system left its parent cluster is called the kinematical age of the system. If the kinematical age is known, it can be used to discriminate between the two scenarios. If an OB-runaway is produced by Blaauw's scenario, the kinematical age and the (current) age of the parent cluster constrain the mass of the progenitor star that exploded during the supernova explosion \citep{Ekstrom2012}.


With a spectral type of O6.5 Iaf+, HD~153919 is the potentially most massive primary in the known sample of HMXBs in the Galaxy \citep[cf.][]{Falanga2015}. The classification f+ refers to the presence of a number of emission lines in the spectrum (He~{\sc ii} 4686, the N~{\sc iii} complex at 4640~\AA, and Si~{\sc iv} 4089,4116) produced by the dense stellar wind \citep{Walborn1990}. The nature of the compact secondary has been subject of a long debate, e.g.\ \citet{BrownGE1996}: it is either a massive neutron star \citep{Clark2002} or a low-mass black hole, with a mass very similar to one of the merging compact objects in GW190412 \citep{Abbott2020}. No X-ray pulsations have been detected that would definitely identify 4U~1700-37 as a neutron star. An observational constraint on the mass ratio of the eclipsing system is obtained from the radial-velocity amplitude $K$ and eccentricity of the orbit of HD~153919 \citep{Hammerschlag2003}.

\cite{Ankay2001} proposed that HD~153919 / 4U~1700-37 originates from the OB association Sco~OB1 of which NGC~6231 is the suggested core (Fig.~\ref{fig:ScoOB1}). Based on {\it Hipparcos} data they derive a kinematical age of $2 \pm 0.5$ million years. They take an upper limit on the age of Sco~OB1 of 8~Myr and conclude that the progenitor of 4U~1700-37 went into supernova within 6~Myr, which puts a lower limit on the initial mass of the progenitor of 30~\(\textup{M}_\odot\). 

NGC~6231 is a young open cluster that hosts many O and B stars and a large pre-main-sequence (PMS) population. It is considered to be the nucleus of the Sco~OB1 association. The star formation process in this cluster has recently ended, and the molecular cloud in which the stars were formed has dispersed. This allows for high quality (optical) observations of the stars in the cluster and for a detailed study of the final outcome of the star formation process. The most recent age determinations of NGC~6231 are listed in Tab.~\ref{tab:6231}.

\begin{table}[h]
	\centering
	\caption{Overview of recent age determinations of NGC 6231. The second column indicates whether the isochrone is fit to the pre-main-sequence (PMS) or main-sequence (MS) population.}
	\begin{tabular}{lll}
		\hline
		\multicolumn{3}{c}{Age determination NGC 6231}                                                     \\ \hline
		\multicolumn{1}{l|}{Age (Myr)} & \multicolumn{1}{|l|}{Isochrone}              & Reference \\ \hline
		1-8                             & PMS                       & \citet{Damiani2016}   \\ 
		1-10                            & PMS                      & \citet{Sana2007}     \\ 
		1-7                             & PMS                      & \citet{Sung2013}      \\ 
		4-7                             & MS              & \citet{Sung2013}      \\ 
		$4.7 \pm 0.4$ & PMS and MS & this work \\ \hline
	\end{tabular}
	\label{tab:6231}
\end{table}

NGC~6231 is the most probable birthplace of the HMXB 4U~1700-37. \citet{Feinstein2003} found evidence for a supernova explosion in NGC 6231 and showed that the observed initial mass function (IMF) allows for the initial presence of one or more very massive stars with masses up to about 70 \(\textup{M}_\odot\) (spectral type O3--O4). \citet{Sung2013} expect that NGC 6231 hosted 2-3 massive stars that were more massive than the currently most massive star in that cluster. 


The astrometric and photometric data provided by the \textit{Gaia} mission \citep{GaiaCollaboration2016, GaiaCollaboration2018a, GaiaCollaboration2020} make it possible to quantitatively confirm the runaway scenario sketched by \citet{Ankay2001} based on {\it Hipparcos} data. The \textit{Gaia} satellite executes an ambitious project of the European Space Agency (ESA) to record astrometric, photometric and spectroscopic data of more than one billion objects in the Galaxy and the Local Group up to $G = 21$ \citep{GaiaCollaboration2016, Riello2018}. This is more than one percent of the stellar population in the Milky Way. With the accurate parallaxes and proper motions, we perform a membership analysis of NGC 6231. Subsequently, its current age is determined and the motion of HD 153919 is obtained relative to NGC~6231.

The scientific importance of confirming this scenario is that the progenitor mass of the compact object in 4U~1700-37 can be determined, the time of the supernova is set, the amount of material lost from the system is constrained, and the time it takes for the (rejuvenated) secondary, the current Of star HD153919, to evolve off the main sequence, can be measured (given the short duration of the HMXB phase). Thus, this provides a unique opportunity to test and constrain evolutionary models of massive binaries yielding binary neutron stars and/or black holes, the progenitors of the recently detected gravitational wave sources \citep[e.g.,][]{Abbott2017}.

In section~\ref{chap:HMXB|sec:membership_ngc6231} we present the membership analysis of NGC~6231. We determine its age in section~\ref{chap:HMXB|sec:age_ngc6231}. We provide an update of the parameters of the HMXB 4U1700-37 in section~\ref{chap:HMXB|sec:update_HD153919}. Section~\ref{chap:HMXB|sec:kinematic_age} presents the detailed reconstruction of the history of the system in space and time. In section~\ref{chap:HMXB|discussion} we discuss different evolutionary scenarios that would fit these observations and address the physical nature of 4U1700-37.

\section{Membership NGC~6231}
\label{chap:HMXB|sec:membership_ngc6231}
The initial sample is manually selected from the {\it Gaia} DR2 database based on astrometric properties: coordinates, proper motion and parallax\footnote{We have completed the membership analysis before \textit{Gaia} Early Data Release 3 (EDR3). We have investigated whether EDR3 would lead to a significantly different outcome regarding cluster membership, age determination, and kinematical age of the system; it does not. However, we used the improved parallaxes and proper motions of EDR3 for the members yielded by the DR2 analysis.}. We started with the candidate members of Sco~OB1 listed in \citet{Ankay2001} to get a first impression of their astrometric properties. Subsequently, we queried the {\it Gaia} database at large in the thus defined parameter space. The resulting sample contains candidate members but also many (unrelated) stars in the field. After applying the astrometric corrections a membership analysis program is used to make a distinction between the field stars and the group members based on their proper motion and parallax \citep[cf.][in prep.]{Guo2021b_prep, Guo2021a_prep}. 

\begin{figure}[h!]
	\centering
	\includegraphics[width=0.8\linewidth]{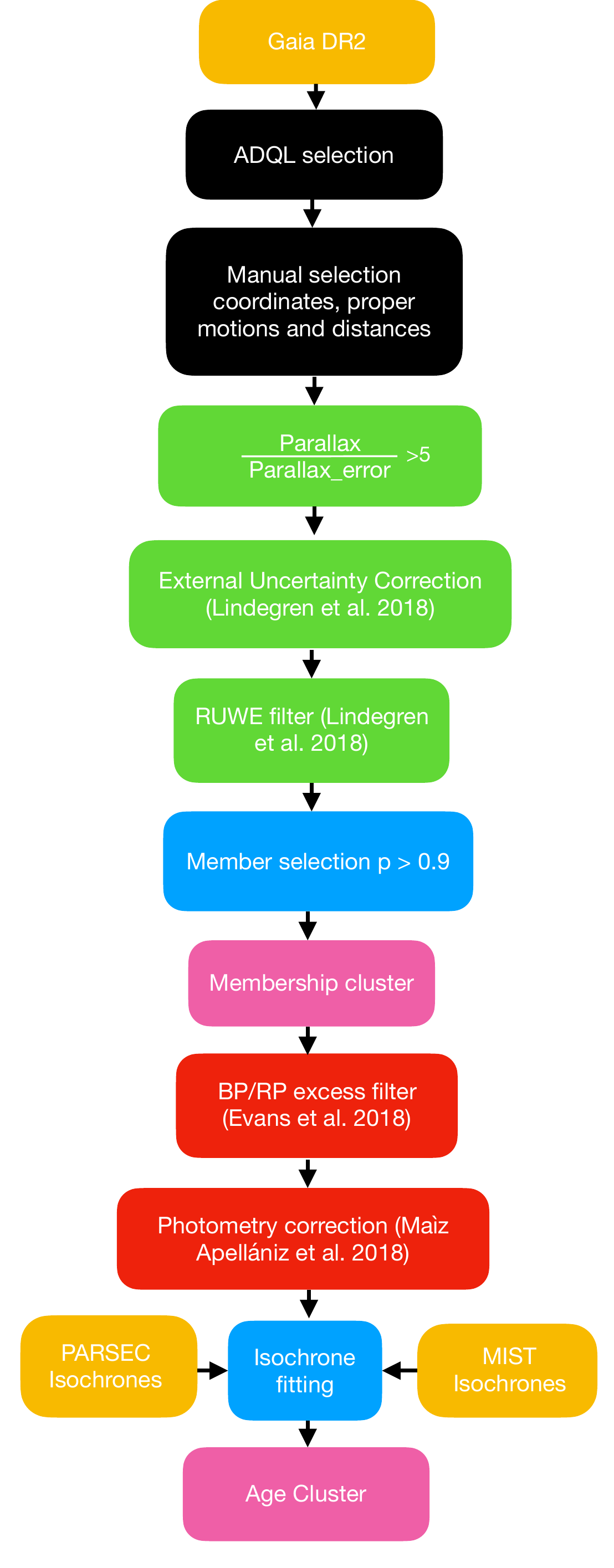}
	\caption{Flowchart showing the scripts that are used to get from the data to the results: yellow blocks refer to data; black blocks represent the applied manual selection criteria; green blocks indicate where astrometric corrections have been applied; red blocks represent the occurrence of photometric corrections; blue blocks signal the pipelines that are used to select the data and pink blocks represent the results.}
	\label{fig:Flowchart}
\end{figure}

\subsection{Manual selection}

We start with a sample downloaded from the \textit{Gaia} archive with an ADQL query (Appendix~\ref{ADQL}). This is done by selecting a range of Galactic longitude, Galactic latitude and parallax based on our prior knowledge of the cluster. The dataset includes astrometry and photometry information of 127,349 sources. The next step is to further reduce the sample based on Galactic longitude, Galactic latitude, parallax and proper motion over-densities using visualization tools such as TOPCAT (\url{http://www.starlink.ac.uk/topcat/}), until we obtain a sample that spans the cluster parameter space, but also includes enough field stars to make a distinction between cluster members and field stars. This is an important prerequisite for the membership analysis program. The remaining sample consists of 9101 sources. After astrometric corrections (Sect.~\ref{sec:astro} and Fig.~\ref{fig:Flowchart}), the remaining sample consists of 3661 sources. The positions and proper motions of the selected stars are shown in Figs.~\ref{fig:coord} and \ref{fig:propa}. A cut was made in distance such that only stars with a distance between 1300~pc and 1900~pc were included in the membership analysis.

\subsection{Corrections to the \textit{Gaia} DR2 data}

In order to use the astrometric and photometric data provided by \textit{Gaia} DR2 some corrections have to be made to the data obtained from the archive. Fig.~\ref{fig:Flowchart} shows all the necessary corrections that are needed to work with the data: astrometric corrections are displayed as green blocks while photometric corrections are shown as red blocks.

\subsubsection{Astrometric corrections}
\label{sec:astro}
After the release of \textit{Gaia} DR2, errors on astrometric variables were found to be underestimated and some sources did not have reliable astrometric data at all. To correct for this all the errors on parallax and proper motion have to be recalculated and some sources have to be discarded from the sample.

\paragraph{External uncertainty correction}

For the Tycho-\textit{Gaia} Astrometric Solution (TGAS) of \textit{Gaia} DR1 the published errors were obtained from a comparison with {\it Hipparcos} parallaxes. No such calibration has been done for \textit{Gaia} DR2. This results in underestimated errors for all sources in \textit{Gaia} DR2; faint sources ($G > 13$) being more affected than bright sources. \citet{Lindegren2018} tested the correctness of the errors provided by the \textit{Gaia} catalog and came up with corrections on the errors of the parallaxes and the proper motions for bright sources ($G < 13$) and faint sources  by comparing astrometric data from other catalogs, such as {\it Hipparcos} data and quasar data, to the astrometric data provided by \textit{Gaia} DR2. The corrections that should be applied are shown in Eq.~\ref{eqn:error} in which $\sigma_{\rm{ext}}$ is the total error, $\sigma_i$ is the error from the \textit{Gaia} catalog and $\sigma_s$ is a value computed by \citet{Lindegren2018} shown in Tab.~\ref{tab:error}. The value of $k$ is 1.08 for all corrections.

\begin{equation}
	\label{eqn:error}
	\sigma_{\rm{ext}} = \sqrt{k^2\sigma^2_i+\sigma_s^2}.
\end{equation}

\begin{table}[h!]
	\caption{Values of $\sigma_s$ needed to calculate the correct errors with Eq.~\ref{eqn:error}.}
	\centering
	\begin{tabular}{l|lll}
		\hline
		& G \textless 13 & G \textgreater 13 & Unit   \\ \hline
		parallax      & $\sigma_s$ = 0.021          & $\sigma_s$ = 0.043             & mas    \\ \hline
		proper motion & $\sigma_s$ = 0.032          & $\sigma_s$ = 0.066             & mas/yr\\
		\hline
	\end{tabular}
	
	\label{tab:error}
\end{table}

\paragraph{The Renormalised Unit Weight Error (RUWE)}

This is a recommended goodness of fit indicator for \textit{Gaia} DR2 astrometry that was not directly listed in the \textit{Gaia} Archive\footnote{Now the archive provides an additional table for RUWE.}. It is a renormalization of the Unit Weight Error (UWE) by $u_0(G,C)$: an empirical normalization factor dependent on photometric properties provided by ESA. If for a source RUWE $>$ 1.4 holds, the goodness of fit is insufficient and it is advised to discard this source from the sample. The parameters needed to calculate UWE and tables with the correct normalization factor $u_0(G,C)$ can be found on, respectively, the \textit{Gaia} archive and the {\it Gaia} DR2 known issues web page\footnote{\url{https://www.cosmos.esa.int/web/gaia/dr2-known-issues}}. The parameters that were used from the \textit{Gaia} archive are the following:

\begin{itemize}
	\item  $\chi^2$ = \texttt{astrometric\_chi2\_al}
	\item N = \texttt{astrometric\_n\_good\_obs\_al}
	\item G = \texttt{phot\_g\_mean\_mag}
	\item C = \texttt{bp\_rp} (if available)
\end{itemize}

\noindent UWE and RUWE are calculated using Eqs.~\ref{eqn:Uwe} and \ref{eqn:Ruwe}:

\begin{equation}
	\label{eqn:Uwe}
	UWE = \sqrt{\chi^2/(N-5)},
\end{equation}

\begin{equation}
	\label{eqn:Ruwe}
	RUWE = UWE/u_0(G,C).
\end{equation}

\noindent No sources in our sample had to be discarded because they all have RUWE $<$ 1.4.

\paragraph{Parallax and distance}


The parallax is essential to determine the distance of the cluster as well as the runaway system, with which we are able to confirm whether the runaway system originates from the parent cluster in 3-dimensional space. However, despite \textit{Gaia}'s unprecedented astrometric accuracy, the parallax measurements for these distant sources are relatively uncertain. Besides, the task of inferring distance from parallax with a properly defined uncertainty is not trivial and is dependent on the choice of a prior function on the distance distribution \citep{Bailer-Jones2018}. The naive method of simply inverting the parallax for distance is only acceptable for \texttt{parallax\_over\_error} $> 5$ \citep[][]{Bailer-Jones2015}. In this work, for simplicity, we use sources with \texttt{parallax\_over\_error} $> 5$ only, and invert them for the distance estimate (and subsequently updated to the EDR3 parallax).

\subsubsection{Photometric corrections}

The measured magnitudes in the \textit{Gaia} $G_{\rm BP}$ and $G_{RP}$ band may be contaminated such that these sources must be discarded from the sample. The $G$-band magnitude is corrected for all the remaining sources because of a systematic error in that band.

\begin{figure}[h!]
	\centering
	\includegraphics[width=\linewidth]{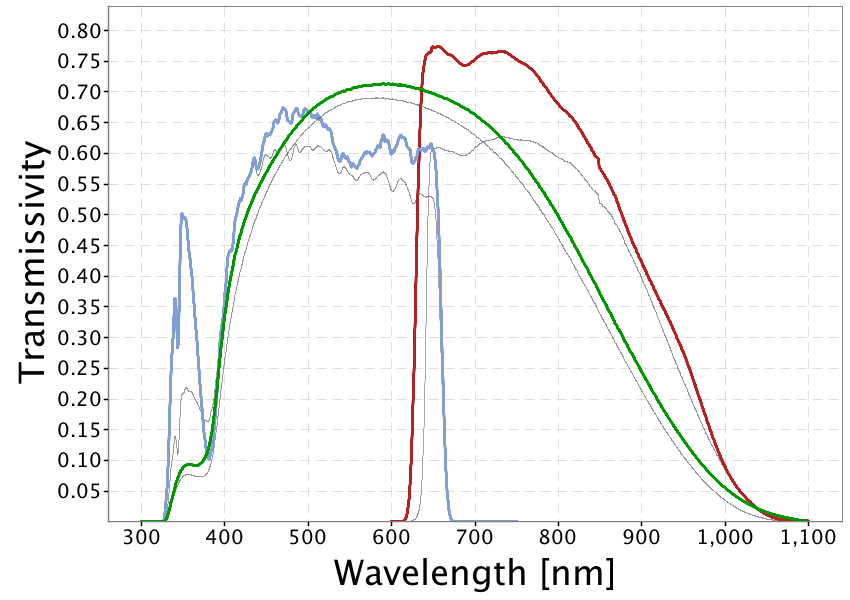}
	\caption{The colored lines show the passbands for $G$ (green), $G_{\rm BP}$ (blue) and $G_{\rm RP}$ (red), defining the \textit{Gaia} DR2 photometric system. The thin, gray lines show the nominal, prelaunch passbands model published in \citet{Jordi2010}, used for \textit{Gaia} DR1. Source: ESA}
	\label{fig:passbands}
\end{figure}

\paragraph{BP/RP excess filter}

For the majority of the sources in the \textit{Gaia} archive the flux is obtained in three bands: $G$, $G_{\rm BP}$ and $G_{\rm RP}$. The flux in the G band is the most reliable because it is determined from profile-fitting to a narrow image. The $G_{BP}$ and $G_{RP}$ bands on the other hand give the total flux of the area of 7.35 $\rm{arcsec}^2$ around the source which makes them more sensitive to pollution from the background or a nearby star. The $G_{BP}$ and $G_{RP}$ bands together cover about the same wavelength range and have similar transmission as the $G$ band. The wavelength range that these passbands span and their transmission is shown in Fig.~\ref{fig:passbands}.

\begin{figure}[h!]
	\centering\includegraphics[width=\linewidth]{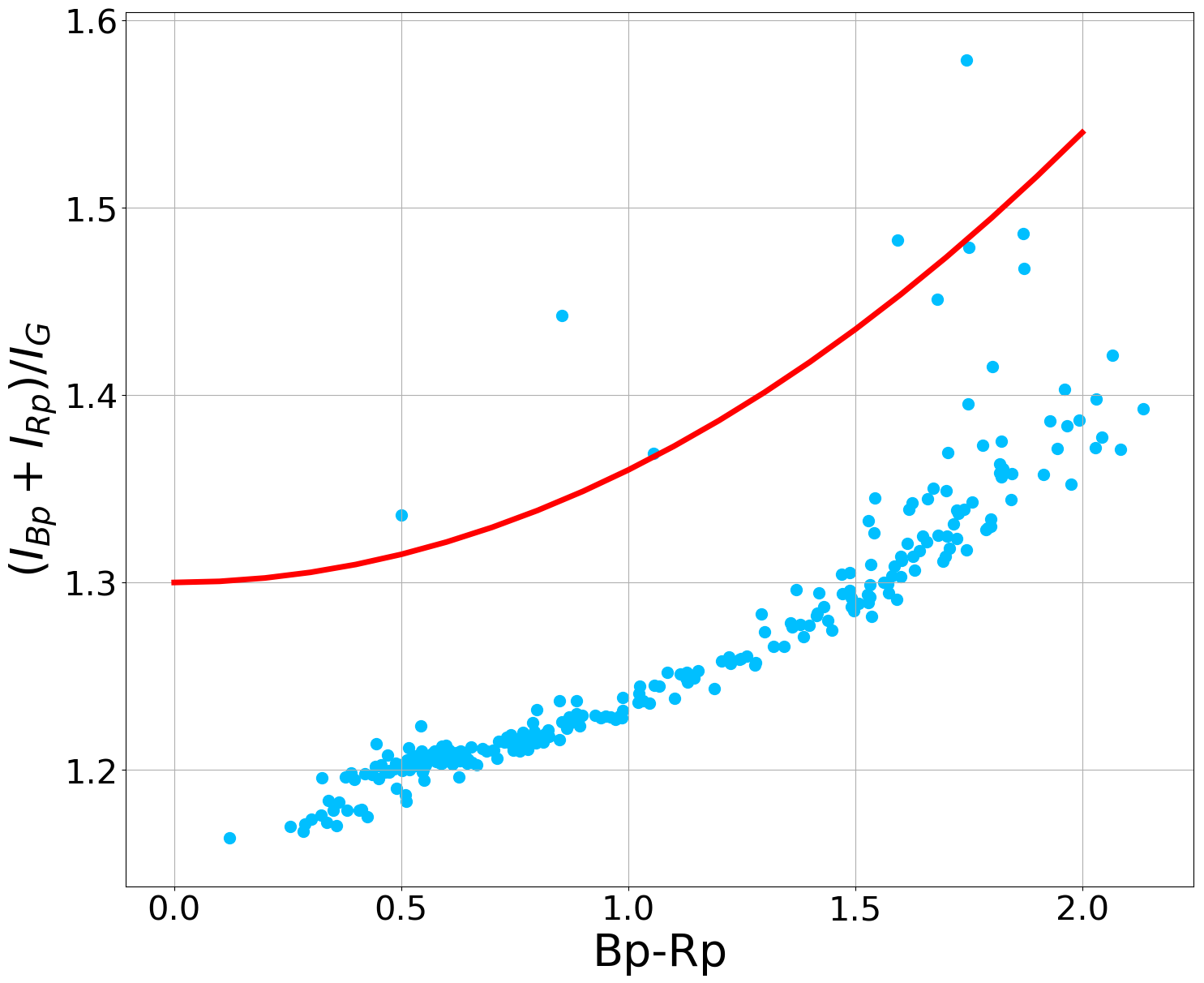}
	\caption{Flux excess versus color of sources for the members of NGC 6231 that were selected using our membership probability code. The red line corresponds to $1.3 + 0.06 \left( G_{\rm BP} - G_{\rm RP} \right)^2$. The uncontaminated sources are below this curve \citep{Evans2018}. Five of the members of NGC 6231 are situated above the red curve; these sources were discarded from the sample for the age determination.}
	\label{fig:BPRP}
\end{figure}

\citet{Evans2018} describe how to measure the pollution in the $G_{\rm BP}$ and $G_{\rm RP}$ bands and how to exclude bad measurements. First the assumption is made that if the $G_{\rm BP}$ and $G_{\rm RP}$ bands are not heavily polluted, the ratio of fluxes $(I_{G_{\rm BP}} + I_{G_{\rm RP}})/I_G$, where $I$ stands for flux, should be slightly bigger than one because the $G_{\rm BP}$ and $G_{\rm RP}$ bands have some overlap and a better transmission at some wavelengths, but together they cover almost the same area as the $G$ band (Fig.~\ref{fig:passbands}). This ratio can be found in the \textit{Gaia} archive as \texttt{phot\_bp\_rp\_excess\_factor}. If the \texttt{phot\_bp\_rp\_excess\_factor} significantly exceeds 1, then there must be pollution in the $G_{\rm BP}$ and/or $G_{\rm RP}$ band. A plot of $(I_{G_{\rm BP}} + I_{G_{\rm RP}})/I_G$ versus $G_{\rm BP} - G_{\rm RP}$ clearly shows that some sources are polluted in the $G_{\rm BP}$ and $G_{\rm RP}$ band. \citet{Evans2018} computed a threshold that distinguishes the "well-behaved" sources from the polluted ones. If for a source: $(I_{G_{\rm BP}} + I_{G_{\rm RP}}) / I_G < 1.3 + 0.06 \left( G_{\rm BP} - G_{\rm RP} \right)^2$ then it will be found under the red curve in Fig.~\ref{fig:BPRP} which means that the $G_{\rm BP}$ and $G_{\rm RP}$ band magnitude can be used. If it is situated above this curve then the $G_{\rm BP}$ and $G_{\rm RP}$ bands of this source are contaminated; this is the case for five sources in our sample.

\paragraph{Correction of the $G$-band magnitude}

The \textit{Gaia} DR2 photometry in the $G$ band is affected by systematic errors. It shows an approximately linear trend between $G = 16$ and $G = 6$ of 3.2 $\pm$ 0.3 mmag/mag \citep{Apellaniz2018}. All the sources in the sample used for this study are in the range 6 $<$ G $<$ 16, and are corrected using:

\begin{equation}
	G^{'} = G - 0.0032(G-6) \textrm{ \hspace{0.1cm} for \hspace{0.1cm} } 6 < G < 16.
\end{equation}

\subsection{Membership probability}

To distinguish cluster members from field stars a program is used to calculate the probability of a star being a member of a moving group based on a maximum-likelihood method which is a modified version of the method described by \citet{Lindegren2000}. It makes use of a kinematic model.

\subsubsection{The velocity model}
\label{velo}

\noindent First we assume that all stellar members of the cluster are moving in the same direction with the velocity vector:

\begin{equation}
	\vec{v}_g = (v_{x,g}, v_{y,g}, v_{z,g})^T.
\end{equation}

\noindent
We use ICRS coordinates in the calculation of the membership analysis. The next step is to use the normal triad $\hat{p}, \hat{q}, \hat{r}$ to project the velocity vector $\vec{v}_g$ on the sky for a given sky coordinate $(\alpha, \delta)$ and parallax $\varpi$ to obtain a proper motion and radial velocity. The components of the normal triad are defined as:

\begin{equation}
	\label{eq:normatriad}
	\hat{p} = \begin{bmatrix}
		-\sin\alpha\\\cos\alpha\\0
	\end{bmatrix}, \ 
	\hat{q} = \begin{bmatrix}
		-\sin\delta\cos\alpha\\-\sin\delta\sin\alpha\\\cos\delta
	\end{bmatrix}, \ 
	\hat{r} = \begin{bmatrix}
		\cos\delta\cos\alpha\\\cos\delta\sin\alpha\\\sin\delta
	\end{bmatrix} \, .
\end{equation}

\noindent
Subsequently, we obtain the proper motion and radial velocity using the normal triad and the group motion:

\begin{equation}
	\label{eq:pmfield}
	\begin{bmatrix}
		\mu_{\alpha*,g}\\\mu_{\delta,g}\\v_{r,g}\\
	\end{bmatrix}
	= 
	\begin{bmatrix}
		p_{x}\varpi/A &p_{y}\varpi/A &p_{z}\varpi/A \\
		q_{x}\varpi/A &q_{y}\varpi/A &q_{z}\varpi/A \\
		r_{x} &r_{y} &r_{z} \\
	\end{bmatrix}
	\cdot 
	\begin{bmatrix}
		v_{x,g} \\ v_{y,g} \\ v_{z,g}
	\end{bmatrix} \, ,
\end{equation}

\noindent
where $\mathrm{A} = 4.74047\,\mathrm{yr\,km\,s^{-1}}$ is the astronomical unit in $\mathrm{yr\,km\,s^{-1}}$, converting the angular unit of proper motion (mas~yr$^{-1}$) to linear velocity (km~s$^{-1}$). We assume a velocity dispersion model $S$ that is the same across the whole group:

\begin{equation}
	{S}_g = 
	\begin{bmatrix}
		\sigma_{x,g}^2 	&0			&0	 		\\
		0			&\sigma_{y,g}^2 &0	 		\\
		0			&0			&\sigma_{z,g}^2
	\end{bmatrix} \, ,
\end{equation}

\noindent
where the subscript $g$ means group. Another assumption is that the dispersions in different directions are uncorrelated and thus we set all the terms except for the diagonal terms to zero. The field model is a copy of the group model. The only difference is the assumption that is made for the dispersion. We assume a small dispersion for the group model and a large dispersion for the field model.

\subsubsection{The likelihood function}

The next step is to apply the kinematic modeling method from \citet{Lindegren2000}. The proper motion of each star $i$, $\vec{\mu_i} = (\mu_{\alpha*, i}, \mu_{\delta, i})^T$ will be compared to both model predictions $\vec{\mu_{i, g}}$ and $\vec{\mu_{i, f}}$: the group proper motion and the field proper motion on the position of the star, respectively, where we assume that the differences $\Delta\vec{\mu_{i,k}} = \vec{\mu_i} - \vec{\mu_k}$, in which $k \in \{g, f\}$, follows a Gaussian distribution. The likelihood function is:

\begin{equation}
	p_{i,k}(\vec{\mu}_k|\vec{\mu}_i) = 
	\frac{1}{2\pi|{D}_i|}
	\exp(\Delta\vec{\mu}_{i,k}^T D_i^{-1} \vec{\mu}_{i,k} / 2),
\end{equation}

\noindent 
in which $D_i$ is the variance matrix:

\begin{equation}
	\begin{aligned}[align=left]
		{D}_i = {C}_i &+ 
		\begin{bmatrix}
			\hat{p}_i^TS\hat{p}_i	&\hat{p}_i^TS\hat{q}_i\\
			\hat{q}_i^TS\hat{p}_i	&\hat{q}_i^TS\hat{q}_i
		\end{bmatrix} \left(\frac{\varpi_i}{\mathrm{A}}\right)^2	\\
		&+
		\begin{bmatrix}
			(\hat{p}_i\cdot\vec{v_k})^2	&(\hat{p}_i\cdot\vec{v_k})(\hat{q}_i\cdot\vec{v_k})	\\
			(\hat{p}_i\cdot\vec{v_k})(\hat{q}_i\cdot\vec{v_k})	&(\hat{q}_i\cdot\vec{v_k})^2
		\end{bmatrix} \left(\frac{\sigma_{\varpi,i}}{\mathrm{A}}\right)^2 \, .
	\end{aligned}
\end{equation}

\noindent 
It consists of three terms: the first term is the covariance matrix of the observed proper motion, the second term is the variance of proper motion caused by the velocity dispersion, and the third term is the variance caused by the uncertainty in parallax.

\noindent 
The total likelihood of the group and field model that is consistent with the proper motion  of the $i$-th star is:

\begin{equation}
	\Phi_i = \lambda_g \cdot p_{i,g}(\vec{\mu}_g|\vec{\mu}_i)+ 
	\lambda_f \cdot p_{i,f}(\vec{\mu}_f|\vec{\mu}_i) \, .
\end{equation} \\

\noindent 
The combination factors ($\lambda_g, \lambda_f$) satisfy $\lambda_g + \lambda_f = 1$. The program numerically maximizes the value of $\prod_{i=1}^N\Phi_i$ for N sources by optimizing the parameters ($\vec{v}_g$, $S_g$, $\vec{v}_f$, ${S}_f$, $\lambda_g$, $\lambda_f$).
When the optimization is finished, the probability that the $i$-th star belongs to a group can be calculated by using the total likelihood as a normalization factor:

\begin{equation}
	p_{i,g} = \frac{\lambda_g \cdot p_{i,g}(\vec{\mu}_g|\vec{\mu}_i)}{\Phi_i} \, .
\end{equation} \\

\noindent 
Logically, the probability that the $i$-th star belongs to the field is:

\begin{equation}
	p_{i,f} = \frac{\lambda_f \cdot p_{i,f}(\vec{\mu}_f|\vec{\mu}_i)}{\Phi_i} = 1 - p_{i,g} \, .
\end{equation}\\

\noindent
Formally $p_{i,g} > 0.5$ is the lowest requirement for a candidate to be considered a member, however, one would use a threshold higher than 0.5 to achieve a clean sample in the color-magnitude diagram for isochrone fitting. In this work a source is considered a cluster member if $p_{i,g} > 0.9$.

\subsection{The members of NGC 6231}

\begin{figure}[h!]
	\centering
	\includegraphics[width=\linewidth]{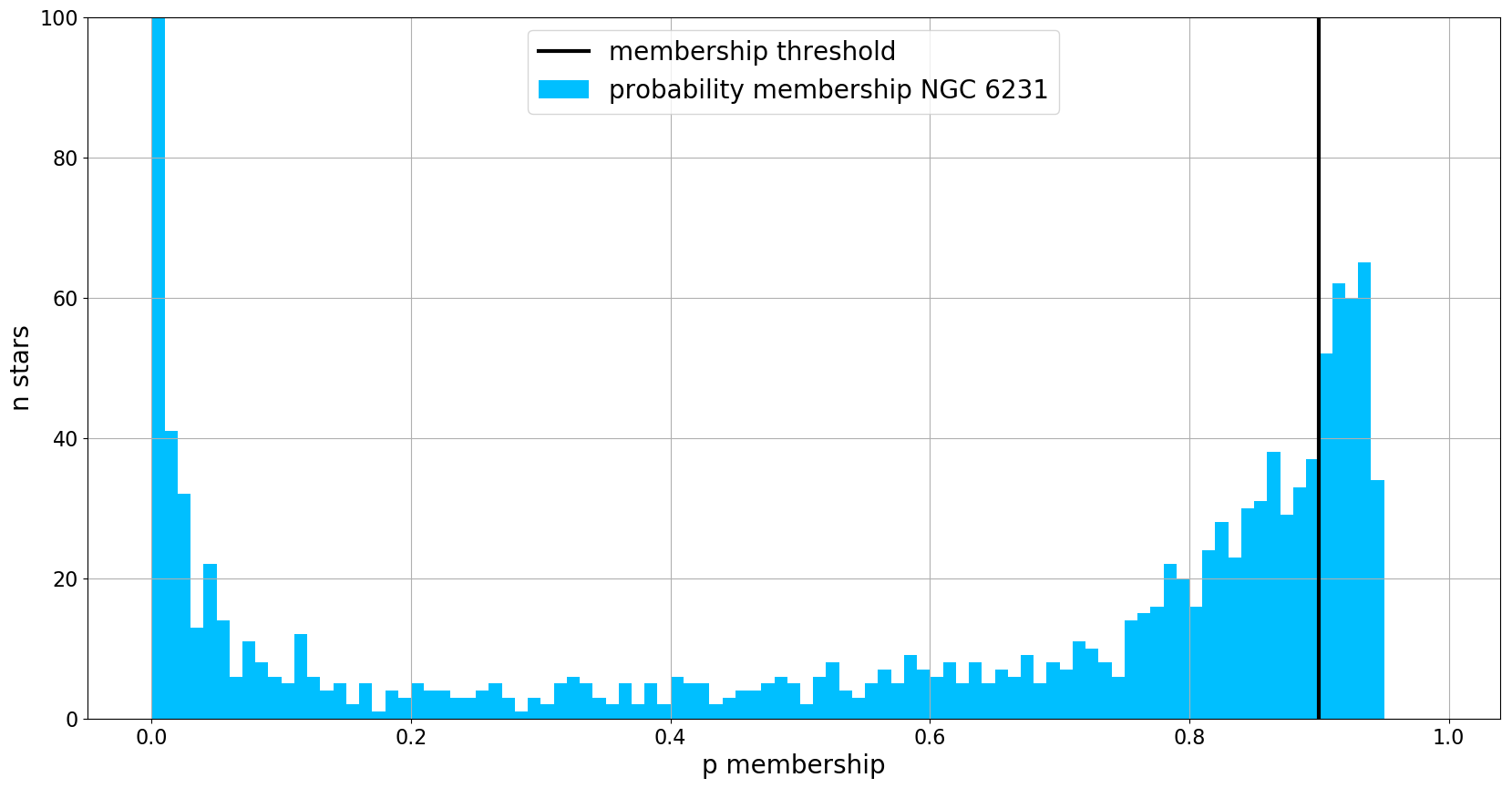}
	\caption{Distribution of membership probability for the stars in NGC 6231 of all the candidates that were assessed by the membership analysis program. The first bin, containing over 2500 stars, is cut off at 100 stars. The bi-modal distribution shows that the program can distinguish members from field stars clearly, leaving only a small amount of ambiguous candidates in the middle range of probability. The black vertical line shows the threshold probability for membership.}
	\label{fig:probs}
\end{figure}

273 stars in NGC 6231 are identified as members with a probability higher than 90\%. The distribution of the membership probability of all candidates is shown in Fig.~\ref{fig:probs}. Although a lower probability threshold would add more stars to the sample, it reduces the accuracy of the isochrone fitting. For more information on how the number of candidates changes with the probability threshold, see Tab.~\ref{tab:pmin_n_relation} in Appendix~\ref{appendix:pmin}.

\paragraph{Update to \textit{Gaia} EDR3} In the light of \textit{Gaia}'s Early Data Release 3 \citep[\textit{Gaia} EDR3, ][]{GaiaCollaboration2020}, which provides a significant improvement on the accuracy of parallax compared to \textit{Gaia} DR2, we update the parallax and proper motions of the members in NGC~6231 and HD 153919/4U 1700-37 without rerunning the membership analysis based on \textit{Gaia} DR2. 

\begin{figure}[h!]
	\centering
	\includegraphics[width=\linewidth]{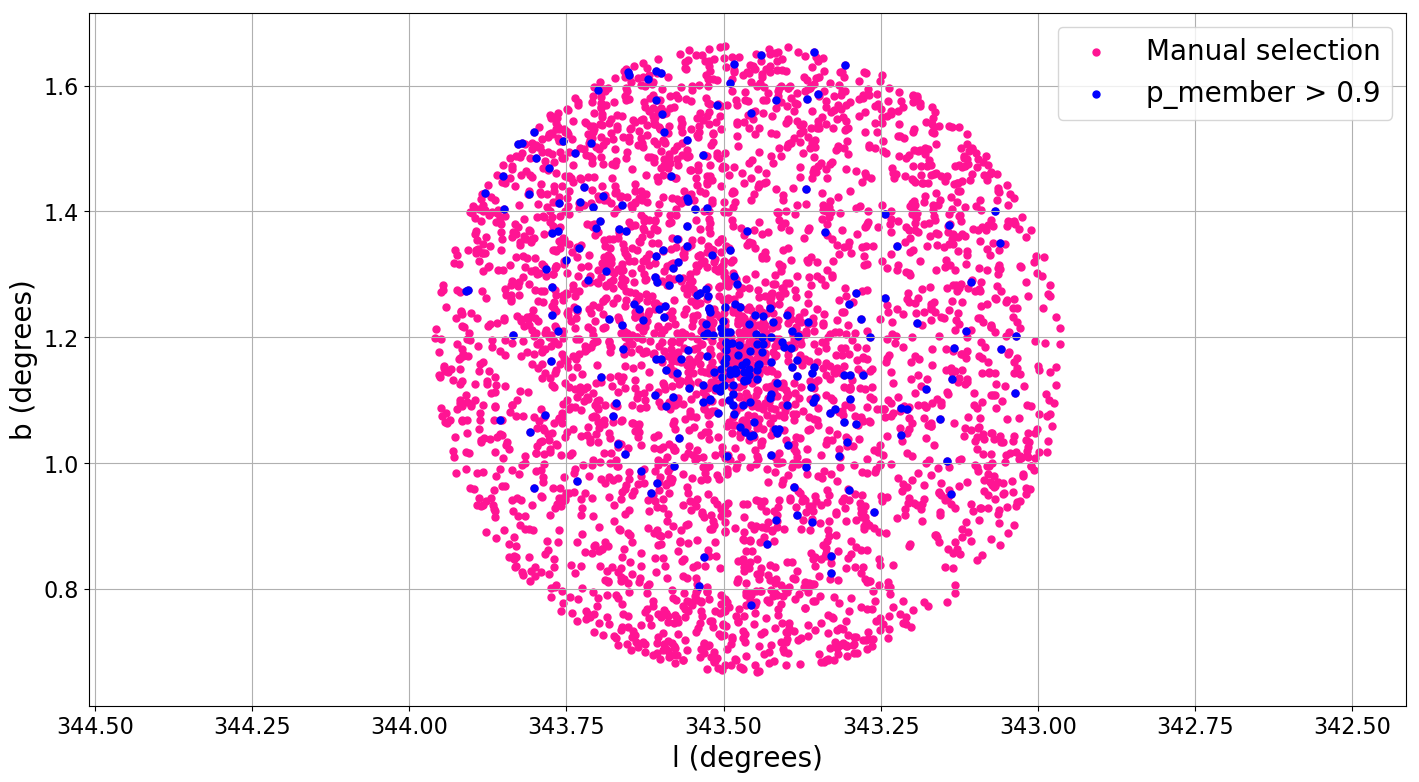}
	\caption{The studied sample projected on the sky (magenta points) and the sources that were selected as members (blue points).}
	\label{fig:coord}
\end{figure}

\begin{figure}[h!]
	\centering
	\includegraphics[width=\linewidth]{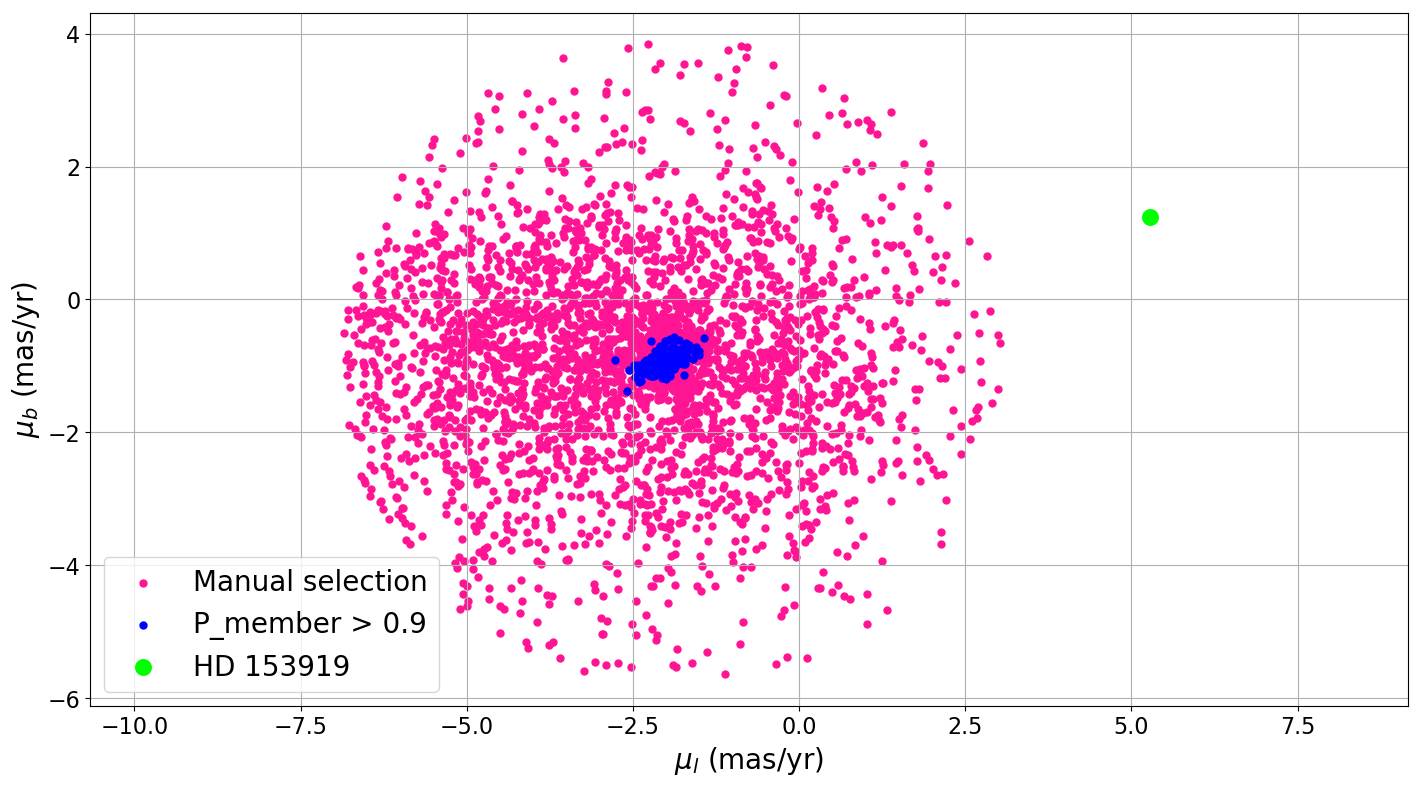}
	\caption{The studied sample in proper motion space (magenta points) and the sources that were selected as members (blue points). One can clearly see that the motion of the members is well confined. The proper motion of HD 153919 (green dot), a clear outlier from the cluster.}
	\label{fig:propa}
\end{figure}

All the stars in the studied sample (magenta), and the selected members (blue) are shown in Fig.~\ref{fig:coord}. The proper motions are displayed in Fig.~\ref{fig:propa}. The clustering of the members in these figures clearly shows that the members are comoving in space.

\begin{figure}[h!]
	\centering
	\includegraphics[width=\linewidth]{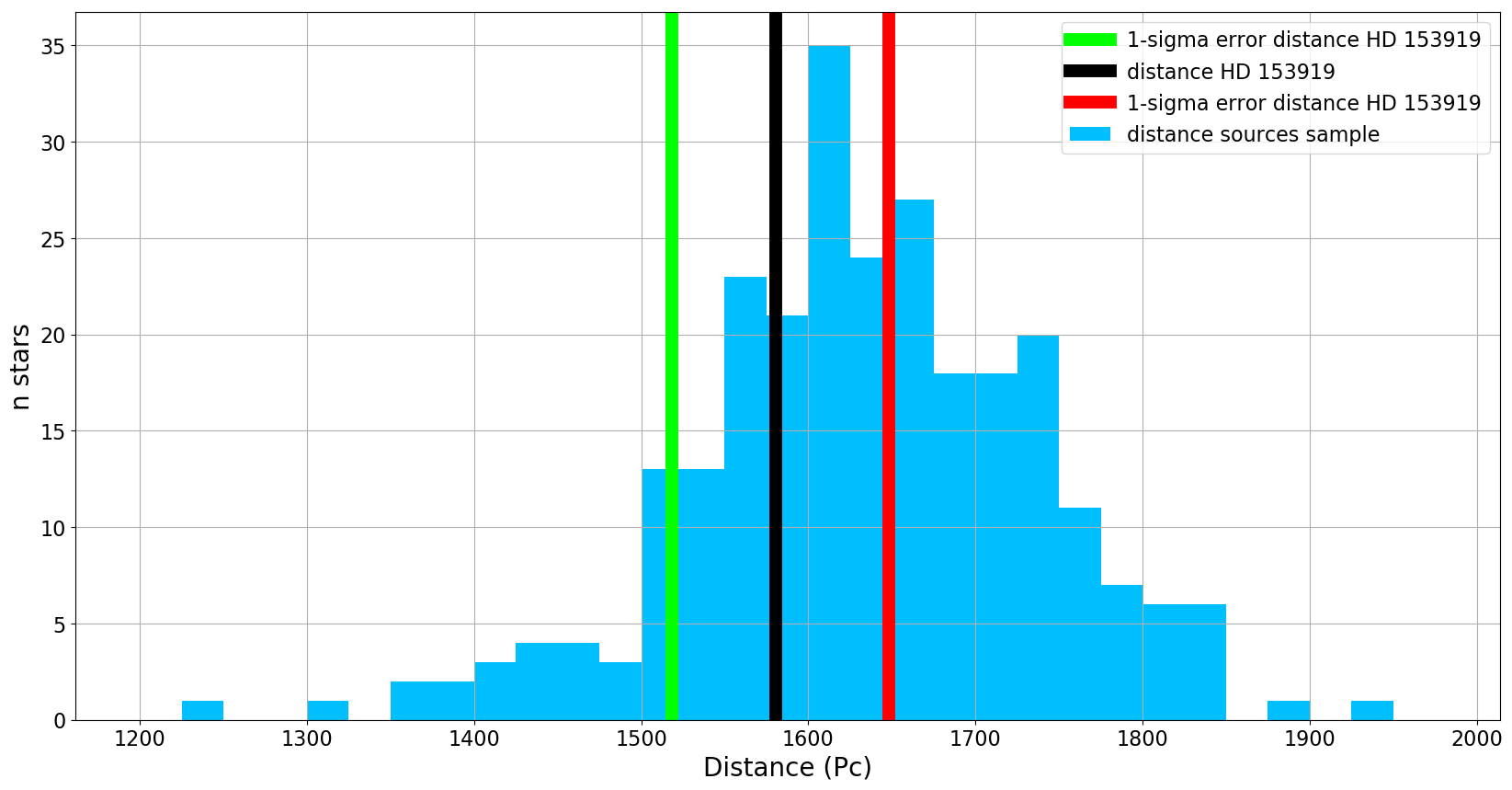}
	\caption{The distance of the members of NGC 6231 (blue histogram) and the most probable distance of HD 153919 (1.58~kpc, black line) with the corresponding distance interval (red and green lines). The average {\it Gaia} EDR3 distance to the cluster is 1.63~kpc, similar to the value determined by \citet{Sana2006} (1.64~kpc) and \citet{Sung2013} (1.59~kpc), both determined with the so-called spectroscopic parallax method (based on the spectral type of the star and the reddening).}
	\label{fig:dista}
\end{figure}

We estimate the distance by inverting the \textit{Gaia} EDR3 parallax for the members in NGC 6231 as well as HD 153919, and plot the histogram in Fig.~\ref{fig:dista}. The inverted mean parallax is at 1.63~kpc, in good correspondence with the distance of 1.64 and 1.59~kpc determined by \citet{Sana2006} and \citet{Sung2013}, respectively. HD 153919/4U 1700-37 is also shown in Fig.~\ref{fig:dista} and appears to be at about the same distance as NGC~6231. In this work, we adopt $1.63 \pm 0.15$~kpc for the distance to NGC~6231.

\section{The age of NGC 6231}
\label{chap:HMXB|sec:age_ngc6231}
With the cluster membership being established, we apply a photometric filter to the members. Five sources have too much pollution in the $G_{BP}$ and $G_{RP}$ bands and are thus discarded. Subsequently, the age determination is performed on 268 of the 273 members of NGC 6231.
The determination of the cluster age was conducted by fitting isochrones to the members of the cluster in the $(G - G_{RP}), G$ color-magnitude diagram. The method that was used is a simplified version of the isochrone fitting method by \citet{Joergensen2005}. We use the code implemented by \citet[in prep.]{Guo2021b_prep}.

\subsection{Isochrone fitting}
\label{isomodel}

The age of NGC 6231 is determined with two different sets of isochrones (PARSEC and MIST), in order to check for consistency. The extinction $A_V$ is treated as a free parameter. The \textit{Gaia} catalog provides an estimate of $A_G$, but these values are not reliable and the conversion to the general extinction $A_{\lambda}$ is not trivial \citep{Jordi2010}. The color excess $E(B-V)$ varies with sight line, and thus the value for $A_V$.
\citet{Sung2013} measured the total-to-selective extinction parameter $R_V$ = 3.22 along this Galactic line of sight and $E(B-V)$ varying between 0.45 and 0.60, with a slightly higher $E(B-V)$ for some of the stars. The authors find an average $E(B-V)$ of 0.47. This gives an $A_V$ of 1.44 to 1.92 with an average of 1.50. Their field of view is somewhat different from ours. They observe a window of $0.67\degree$ by $0.67\degree$ while our field of view is a circle with a radius of $0.5 \degree$. Both sets of isochrones include 31 different values of $A_V$ ranging from 0 to 3 mag in steps of 0.1. We adopt the solar metallicity: $\left[{\rm Fe/H} \right]$ = 0.

In this work we fit the isochrones to three sets of samples within the population of the cluster NGC 6231: the full sample, the pre-main-sequence (PMS) population (a subset of the full sample with $(G - G_{RP}) > 0.65$), and the main-sequence (MS) population (the complementary subset with $(G - G_{RP}) < 0.65$), so that the potential variation in age by analyzing different subpopulations is monitored. We do not take binarity into account.

\subsubsection{PARSEC isochrones}

For the PARSEC isochrones we apply a range from 0.2~Myr to 25~Myr with linear steps of 0.1 Myr. We used PARSEC release v1.2S +  COLIBRI S\_35. All the relevant isochrone parameters can be found in the Appendix in Sect.~\ref{PARSEC} \citep{Bressan2012, Chen2014, Chen2015, Marigo2017, Pastorelli2019, Tang2014}.

\subsubsection{MIST isochrones}
For the MIST isochrones we use the range from 0.2~Myr to 25~Myr with steps of 0.1 Myr. We used MIST version 1.2. All the relevant isochrone parameters can be found in the Appendix in Sect.~\ref{MIST} \citep{Dotter2016, Choi2016, Paxton2011, Paxton2013, Paxton2015}.

\subsection{Likelihood functions}
\label{sec:iso}

On the color--absolute magnitude diagram (CMD) a 2-dimensional Gaussian function can be computed for every star with index $i$ with color $c_i$, absolute magnitude $m_i$, and their errors $\sigma_{c_i}$, $\sigma_{m_i}$:

\begin{equation}
	g_i(c, m) = \frac{1}{2 \pi \sigma_{c_i} \sigma_{m_i}} \exp\left\{-\frac{1}{2} 
	\left[\left(\frac{c - c_i}{\sigma_{c_i}}\right)^2 + 
	\left(\frac{m - m_i}{\sigma_{m_i}}\right)^2\right]\right\} \, .
\end{equation}\\

\noindent 
Every point ($c$, $m$) on the CMD can be evaluated by $g_i(c, m)$ and returns a probability of that point being the true color and magnitude of the star.
Each isochrone of age $a$ adjusted for extinction $\epsilon$ can be described as a curve $I_{a, \epsilon}(c, m)$. Now we can evaluate for every star $i$ the probability for each isochrone $I_{a, \epsilon}(c, m)$ representing the age of the star

\begin{equation}
	G_i(a, \epsilon) = \sum_{j} g_i(c_j, m_j) \, ,
\end{equation}

\noindent
where $j$ represents the index of the points on curve $I_{a, \epsilon}(c, m)$.
Every star $i$ now has a $G$-function $G_i(a, \epsilon)$ with a value for the likelihood for each isochrone $I_{a, \epsilon}(c, m)$. To calculate the age of a single stellar population, if we assume the formation process is coeval, we multiply the G-functions of all the stars to get one likelihood function with an evaluation of each isochrone used for the entire population:

\begin{equation}
	\label{eq:age-ext-dist}
	G (a, \epsilon) = \prod_{i} G_i (a, \epsilon) \, .
\end{equation}

\noindent 
The maximum value of the likelihood function represents the best fit isochrone age and extinction for this cluster.

\subsection{The age of NGC 6231}

To determine the age of the stellar population of NGC~6231, the grid of PARSEC and MIST isochrones was used as described in sections \ref{isomodel} and \ref{sec:iso}. 

\begin{figure}[h!]
	\centering
	\includegraphics[width=\linewidth]{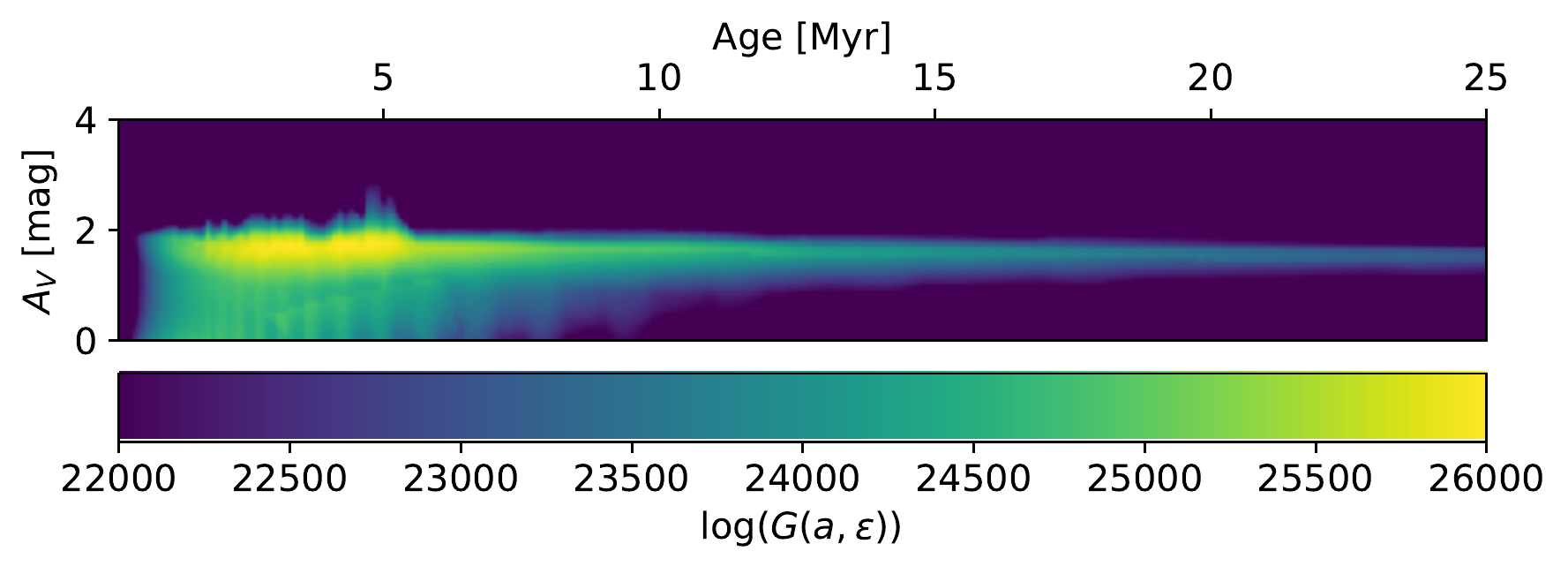}
	\caption{$G$-function distribution in age and extinction based on the PARSEC isochrones for the whole sample. The likelihood is shown with a color scale indicated with the bar at the bottom of the figure. The best fit in this case is the isochrone of 4.7 Myr with an $A_V$ of 1.7 (see also Fig.~\ref{fig:all_iso}).}
	\label{fig:parfull}
\end{figure}

\paragraph{PARSEC isochrone fitting}
The results of the isochrone fitting with the grid of PARSEC isochrones are shown in Fig.~\ref{fig:parfull}. The plot shows a distribution based on the isochrone likelihood evaluation in the age-extinction space; the likelihood values are color coded (scale shown in the color bar). The location of the highest likelihood corresponds to the best fit isochrone of 4.7 Myr with an $A_V$ of 1.7. 

\begin{figure}[h!]
	\centering
	\includegraphics[width=\linewidth]{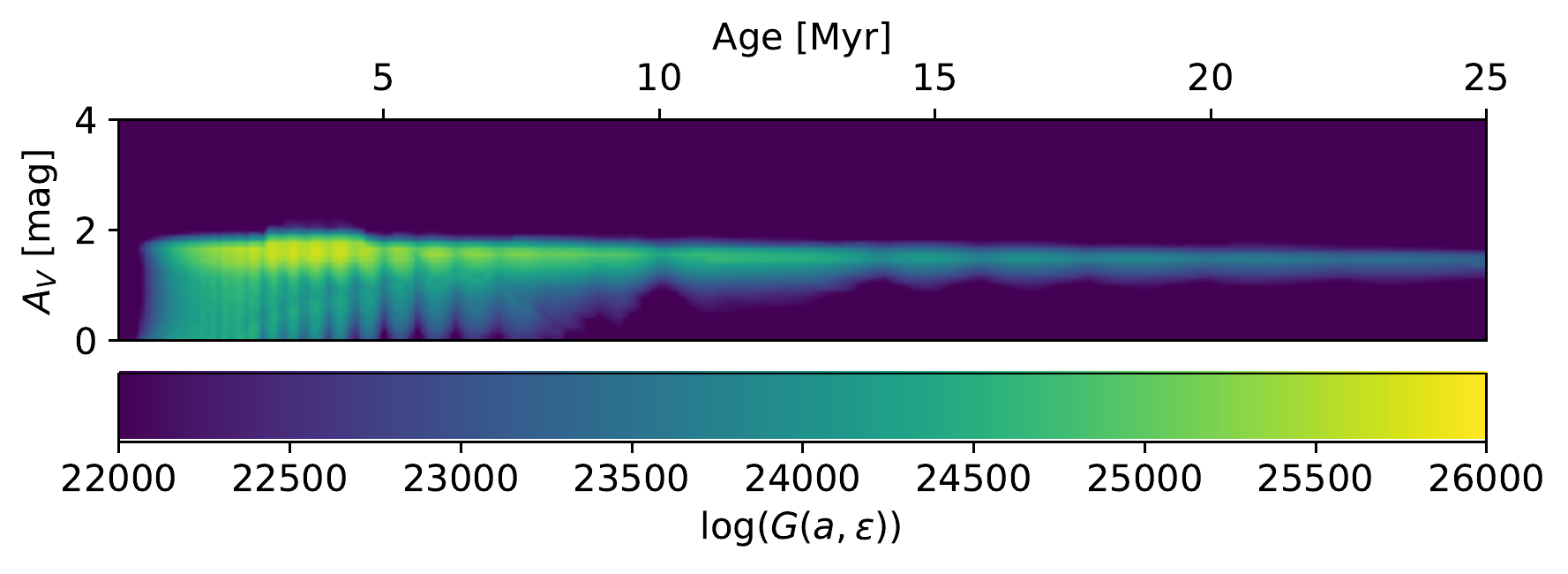}
	\caption{As Fig.~\ref{fig:parfull}: result obtained with the MIST isochrones for the whole sample. The best fit is the isochrone of 3.8 Myr with an $A_V$ of 1.5 (see also Fig.~\ref{fig:all_iso}), with a relatively large range in age.}
	\label{fig:mistfull}
\end{figure}

\paragraph{MIST isochrone fitting}
The results of the $G$-function of MIST isochrone fitting is shown in Fig.~\ref{fig:mistfull}. The outcome is somewhat different from that obtained from the PARSEC isochrones: the best fit is the isochrone of 3.8 Myr with an $A_V$ of 1.5, with a relatively large range in age.

\begin{figure*}[h!]
	\centering
	\includegraphics[width=\textwidth]{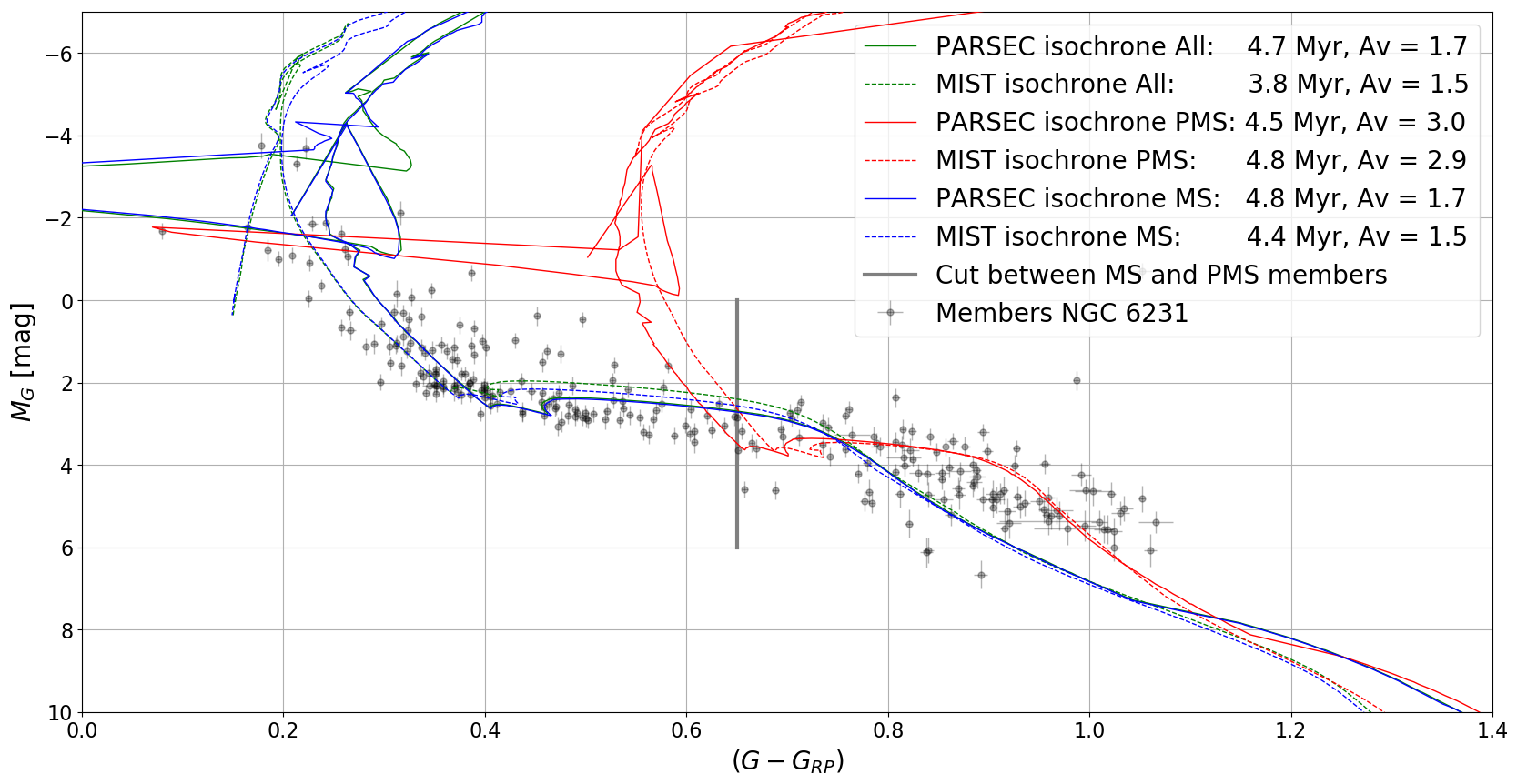}
	\caption{The best fit MIST and PARSEC model isochrones to the full population, the main sequence only, and pre-main sequence only population, respectively. The absolute magnitude $M_{G}$ is based on the EDR3 parallax. It is clear that the isochrones that was fitted to the pre-main sequence only miss the fit due to the absence of the constraint from the MS stars, resulting in inconsistent extinction, regardless of models. Of the 4 remaining isochrones, the range in the age is between 3.8 Myr and 4.8 Myr and the $A_V$ that was determined is around 1.5 -- 1.7 for all models, consistent with the literature values.}
	\label{fig:all_iso}
\end{figure*}

The best fitting isochrones based on both the PARSEC and MIST models are displayed in Fig.~\ref{fig:all_iso}. Both models result in an acceptable fit to the data, however, with a slightly different outcome regarding the age. To investigate this further, we applied the same isochrone fitting procedure to subsamples of the population: (i) the PMS population; (ii) the MS population.

\paragraph{Age determination of the MS population}
The best fit of the PARSEC isochrones to only the MS population of NGC 6231 results in an age of 4.8 Myr with an $A_V$ of 1.7 mag. With the MIST isochrones we find an age of 4.4 Myr with an $A_V$ of 1.5. Both results are similar to the fit on the whole population.

\paragraph{Age determination of the PMS population} 

Fitting the PARSEC and MIST isochrones to only the PMS population gives a similar age result of 4.5 and 4.8~Myr, respectively. However, both fits yield a higher extinction of 3.0 and 2.9 mag in $A_V$, double the literature value and our other results, with an $A_V$ of 1.6. In this case, the best fitting isochrones miss the MS part of the population completely (Fig.~\ref{fig:all_iso}) due to the lack of constraint from the MS sources in the CMD.

\paragraph{Estimating the uncertainties}

We use a bootstrap method to estimate the uncertainty in the age. We repeat the age determination 200 times on randomly selected subsets of the sample. Both sets of isochrones are fit to a randomly selected sample that contains 90\% stars of the original sample. For each sub-sample (All, PMS, MS), the bootstrap method returns the same result in the majority of the 200 trials. The estimated age and extinction of the whole and MS-only populations are consistent with only small standard deviation, while the PMS fittings show less stability with spurious results at very low age and extinction.

\begin{figure}
	\centering
	\includegraphics[width=1.0\linewidth]{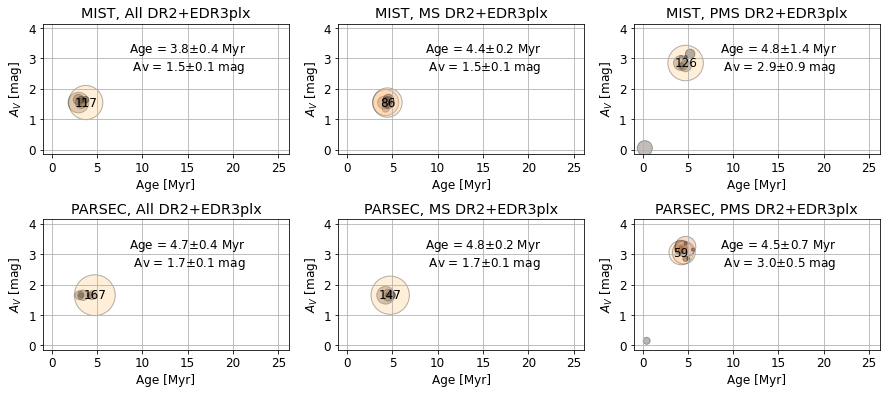}
	\caption{Distribution of solutions based on the bootstrap method. The location of the circle represents the age and extinction of the best fit isochrone, and the size of the cirle is proportional to the occurrence rate based on the 200 trials.}
	\label{fig:age-av-bootstrap-results}
\end{figure}

Table~\ref{tab:age_summary} and Fig.~\ref{fig:age-av-bootstrap-results} show the results of the age and extinction determination with estimated uncertainty. In this work we adopt the results from the PARSEC isochrones fit to the whole sample, as the PARSEC model shows better consistency and stability in the results between the ``all'' and ``MS'' sample than the MIST model. In conclusion, our analysis results in an age and extinction of NGC 6231 of $4.7\pm0.4$ Myr and $A_V = 1.7\pm0.1$ mag, respectively.

\begin{table}[h]
	\centering
	\caption{Age determinations of NGC 6231 based on the {\it Gaia} membership. The PARSEC and MIST models yield similar results on both the MS and whole population of NGC~6231. The fit to the PMS population failed to estimate the extinction and missed the MS population completely due to the lack of constraint in the MS section (see Fig.~{\ref{fig:all_iso}}). The isochrone grid size is 0.1~Myr. In this work, we adopt the result of fitting to the PARSEC isochrone on the whole population as the age and extinction of the cluster.}
	\begin{tabular}{llll}
		\hline                                                  
		Model   & Age (Myr)     & $A_V$ (mag)   & Population  \\
		\hline
		PARSEC  & $4.7 \pm 0.4$           & $1.7 \pm 0.1$           & All          \\
		MIST    & $3.8 \pm 0.4$           & $1.5 \pm 0.1$           & All       \\
		\hline
		PARSEC  & $4.8 \pm 0.2$           & $1.7 \pm 0.1$           & MS           \\
		MIST    & $4.4 \pm 0.2$           & $1.5 \pm 0.1$           & MS           \\
		\hline
		PARSEC  & $4.5 \pm 0.7$           & $3.0 \pm 0.5$           & PMS           \\
		MIST    & $4.8 \pm 1.4$           & $2.9 \pm 0.9$           & PMS           \\
		
		\hline
		{\bf Adopted}     & $4.7\pm0.4$       & $1.7 \pm 0.1$       & All    \\
		\hline 
	\end{tabular}
	\label{tab:age_summary}
\end{table}

\section{Updating the system parameters of HD~153919 / 4U~1700-37}
\label{chap:HMXB|sec:update_HD153919}
The new distance determination of HD~153919, the optical counterpart of the high-mass X-ray binary 4U~1700-37, provides the opportunity to reassess the system parameters. Based on the \textit{Gaia} EDR3 parallax, the distance to the system becomes $1.58 \pm 0.07$~kpc (we note that the DR2 parallax resulted in $1.75 \pm 0.24$~kpc). The effective temperature corresponding to the O6.5~Iaf+ spectral type is $35,000 \pm 1000$~K \citep{Clark2002}. With the X-ray eclipse duration the radius of the O supergiant can be determined: $R_{p} = 25.1 \pm 4.0$~R$_{\odot}$, using an inclination $i = 62\degree \pm1\degree$ and an eclipse semiangle $\theta_E = 32\degree \pm1\degree$ \citep{Falanga2015}. The (black-body) luminosity of HD 153919 becomes $\log(L/{L}_\odot) = 5.93 \pm 0.05$. This is consistent with the (astrometric) luminosity of HD153919, taking into account the photometry, bolometric correction, and extinction.

\citet{Clark2002} used Monte Carlo simulations to determine the mass of HD~153919 and 4U~1700-37: $M_1 = 58$ $  \pm $ $ 11.5 $ \(\textup{M}_\odot\) and $M_X = 2.44$ $  \pm $ $ 0.27 $ \(\textup{M}_\odot\), respectively. The latter mass is relatively high for a neutron star, but low for a stellar-mass black hole. \citet{Clark2002} report that the Monte Carlo simulations fit a zero eccentricity solution and a solution with an eccentricity of 0.22 equally well. The zero eccentricity solution is the conservative approach, because the masses of the O supergiant and compact object would be significantly larger with an eccentric orbit. They state that masses that high do not match the spectral type and $\log{g}$ of the O~supergiant.

\citet{Hammerschlag2003} favor an eccentric orbit of $e$ = 0.22. They state that \citet{Clark2002} miscalculated the masses of the binary members because the value for $K$ (20.6~km~s$^{-1}$) that was used by them was determined using an eccentric orbit; for a circular orbit they should have used $K$ = 18.7~km~s$^{-1}$. If they would have used the correct value of $K$ for the circular orbit, then the masses would have been even higher than those proposed by \citet{Clark2002} for the eccentric solution. \citet{Hammerschlag2003} also found a trend in the residuals of $K$ while trying to fit a circular orbit that could be solved when eccentricity was included. \citet{Hammerschlag2003} determine that the masses of the members of the binary are about 4\% higher than determined by \citet{Clark2002} with $K$ = 20.6 and $e$ = 0.22. They calculated for HD 153919/4U 1700-37 $M_P = 60 \pm 11$ and $M_X = 2.54 \pm 0.27$. With the masses and mass function obtained by \citet{Hammerschlag2003} we find $sin(i) = 0.88~(\pm 0.14)$, $i = 62$. This corresponds well with the value that was determined by \citet{Falanga2015} ($i = 62\degree \pm1\degree$). The latter authors arrive at somewhat lower masses, but we adopt the masses of \citet{Hammerschlag2003}.

A list of the updated  parameters is given in Tab.~\ref{tab:4u}. These parameters are internally consistent and will be used during the remainder of the paper.

\begin{table*}[h!]
	\centering
	\caption{Updated parameters of the system HD 159313/4U1700-37, including error estimates and references.} 
	\begin{tabular}{llll}
		\hline
		\multicolumn{4}{c}{{\bf HD 153919/4U1700-37}}                                                                                        \\ \hline
		\multicolumn{1}{l|}{Quantity}      & \multicolumn{1}{l|}{Value} & \multicolumn{1}{l|}{Error} & \multicolumn{1}{l}{Reference} \\
		\hline
		\multicolumn{1}{l|}{Spectral type}    & O6.5 Iaf+                     &           & \citet{Walborn1973}                    \\ \hline
		\multicolumn{1}{l|}{$T_{\rm{eff}}$ (Kelvin)}  & 35,000                       & $\pm$ 1000                   & \citet{Clark2002}            \\ \hline
		\multicolumn{1}{l|}{$d_{HD~153919}$ (kpc)}     & 1.58                       & $\pm$ 0.07                    & \textit{Gaia} EDR3 \citet{GaiaCollaboration2020}                      \\ \hline
		\multicolumn{1}{l|}{$d_{NGC~6231}$ (kpc)}     & 1.63                       & $\pm$ 0.15                  &  \textit{Gaia} EDR3 \citet{GaiaCollaboration2020}      \\ \hline
		\multicolumn{1}{l|}{log(L/\(\textup{L}_\odot\))}  & 5.93                     & $\pm$ 0.05             &  This work \\ \hline
		\multicolumn{1}{l|}{f(m) (\(\textup{M}_\odot\))}    & 0.0029                     &           & \citet{Hammerschlag2003}                    \\ \hline
		\multicolumn{1}{l|}{P (days)}       & 3.411581                     &    $\pm$ 0.000027          & \citet{Islam2016}                     \\ \hline
		\multicolumn{1}{l|}{K (km/s)}       & 20.6                       & $\pm$ 1.0                       & \citet{Hammerschlag2003}                  \\ \hline
		\multicolumn{1}{l|}{e}              & 0.22                       & $\pm$ 0.04                    & \citet{Hammerschlag2003}                 \\ \hline
		\multicolumn{1}{l|}{$sin(i)$}    & 0.88                         &           $\pm$ 0.14                  &   This work              \\ \hline
		\multicolumn{1}{l|}{$M_X$ (\(\textup{M}_\odot\))}      & 2.54  &    $\pm$ 0.27                   & \citet{Hammerschlag2003}                     \\ \hline
		\multicolumn{1}{l|}{$M_P$ (\(\textup{M}_\odot\))}      & 60             &      $\pm$ 11      & \citet{Hammerschlag2003}                  \\ \hline
		\multicolumn{1}{l|}{a (\(\textup{R}_\odot\))}       & 38                  &      $\pm$ 1        &    Kepler's 3rd law                \\ \hline
		\multicolumn{1}{l|}{$\theta_E$(degrees)}  & $32\degree$             & $\pm$ $1\degree$                  & \citet{Falanga2015}                   \\ \hline
		\multicolumn{1}{l|}{$R_P$ (\(\textup{R}_\odot\))}      & 25.1             &  $\pm$ 4.0   &   This work             \\ \hline
		\multicolumn{1}{l|}{$\dot{M}$ (\(\textup{M}_\odot\)/yr)} & $>2.1 \cdot 10^{-6}$              &          & \citet{Falanga2015}                \\ \hline
		\multicolumn{1}{l|}{$V_\infty$ (km/s)}      & 1700               & $\pm$ 100                     & \citet{VanLoon2001}             \\ \hline
		\multicolumn{1}{l|}{$R_L$ (\(\textup{R}_\odot\))}      & 27.6                       & $\pm$ 0.2       & \citet{Falanga2015}        \\ \hline
		\multicolumn{1}{l|}{$V_r$ (km/s)}    & -64.5                        &           $\pm$  1.5                & \citet{Hutchings1974}                      \\ \hline
		\multicolumn{1}{l|}{$\mu_l$ (mas/yr)}   & 5.46                       &        $\pm$ 0.05                   & \textit{Gaia} EDR3 \citet{GaiaCollaboration2020}                       \\ \hline
		\multicolumn{1}{l|}{$\mu_b$ (mas/yr)}   & 1.13                       &         $\pm$ 0.03                   & \textit{Gaia} EDR3 \citet{GaiaCollaboration2020}                      \\ \hline
	\end{tabular}
	\label{tab:4u}
\end{table*}

\section{The kinematical age of the runaway system}
\label{chap:HMXB|sec:kinematic_age}
\label{sec:kin}

The aim of this section is to confirm that Sco~OB1, and more specifically NGC~6231, is the parent OB association of the HMXB HD~153919 / 4U~1700-37 by reconstructing the path of the system based on {\it Gaia} DR2 data. In this way, the kinematical age of the system, and thus the time of supernova, can be determined. All members of NGC~6231 as well as HD 153919 have well measured proper motions. The radial velocities of the stars in NGC 6231 are not available in \textit{Gaia} DR2, additionally, \citet{Sartoretti2018} state that the radial velocities of stars in DR2 with $T_{\rm{eff}} > 7000$ K are unreliable. Therefore, radial velocities from earlier studies are used in the reconstruction of the path. For simplicity, the Galactic potential has not been taken into account when calculating the path.

\subsection{Reconstruction of the path}

To trace back the paths of the individual stars the coordinates, parallaxes, and proper motions were converted into a Cartesian coordinate system. The transformation of the coordinates is done with the following equations:

\begin{equation}
	x_0 = d\cos{b}\cos{l}
\end{equation}
\begin{equation}
	y_0 = d\cos{b}\sin{l}
\end{equation}
\begin{equation}
	z_0 = d\sin{b},
\end{equation}

\noindent where $d$ is the distance, $l$ is the Galactic longitude and $b$ is the Galactic latitude. Sect.~\ref{velo} describes how to transform vectors from a spherical coordinate system to a Cartesian coordinate system using the normal triad $(\hat{p}, \hat{q}, \hat{r})$, see Eqs. \ref{eq:normatriad} and \ref{eq:pmfield}. To transform vectors from a spherical coordinate system to a Cartesian coordinate system we have to multiply the proper motions and radial velocity with the inverse of the normal triad matrix $(\hat{p}, \hat{q}, \hat{r})^{-1}$. The inverse of matrices representing a spatial rotation operation is the same as the transposed matrix $(\hat{p}, \hat{q}, \hat{r})^T$. The resulting operation is the following:

\begin{equation}
	\begin{bmatrix}
		v_{x} \\ v_{y} \\ v_{z}
	\end{bmatrix}
	= 
	\begin{bmatrix}
		p_{x}dA &q_{x}dA &r_{x} \\
		p_{y}dA &q_{y}dA &r_{y} \\
		p_{z}dA &q_{z}dA &r_{z} \\
	\end{bmatrix}
	\cdot 
	\begin{bmatrix}
		\mu_{\alpha*,g}\\\mu_{\delta,g}\\v_{r,g}\\
	\end{bmatrix}\, ,
\end{equation}

\noindent 
with $d$ the distance and $\mathrm{A} = 4.74047\,\mathrm{yr\,km\,s^{-1}}$ to convert from km~s$^{-1}$ to AU~year$^{-1}$. After transforming the coordinates and velocities from spherical coordinates to Cartesian coordinates it is straightforward to trace the positions of the sources ($x, y, z$) back in time $t$.

\begin{equation}
	x = x_0 + v_x t
\end{equation}
\begin{equation}
	y = y_0 + v_y t
\end{equation}
\begin{equation}
	z = z_0 + v_z t
\end{equation}

\noindent
where ($x_0, y_0, z_0$) is the position of the sources today, and the time $t < 0$ as we move the positions back in time. After obtaining the traced back Cartesian coordinates of the sources one can easily transform them back into the Galactic coordinate system:
\begin{equation}
	l = \tan^{-1}({y/x})
\end{equation}
\begin{equation}
	b = \tan^{-1}({z/\sqrt{x^2 + y^2}})
\end{equation}
\begin{equation}
	d = \sqrt{x^2 + y^2 + z^2}
\end{equation}

\subsection{Radial velocities}

\subsubsection{The radial velocity of NGC 6231}
We obtained the radial velocity of seven OB stars in NGC 6231 from the literature (see Tab.~\ref{tab:radvel}) and used the mean value to represent the radial velocity of the cluster. All these OB~stars have a binary companion except for HD 152314; HD 152270 is a Wolf-Rayet + O star binary. The listed radial velocities of the binaries are based on multiple measurements at different orbital phases of the binaries and the orbital motion is removed. The mean radial velocity of the cluster is $-33.5 \pm 4.0$ km~s$^{-1}$, in agreement with the measurement of $-27.3$ km~s$^{-1}$ by \citet{Kharchenko2005}.

\begin{table*}[h!]
	\centering
	\caption{
		Spectral type, radial velocity, membership probability, and the distance based on \textit{Gaia} EDR3 \citet{GaiaCollaboration2020}, for members of NGC 6231 and HD153919/4U1700-37. The mean radial velocity of the cluster members is $-33.4 \pm 4.0$ km~s$^{-1}$. Notes: [1] \citet{Sana2009}, [2] \citet{Pourbaix2005}, [3] \citet{Hill1974}, [4] \citet{Wilson1953}, [5] \citet{Gies1986}.}
	\begin{tabular}{l|l|llll}
		Name        & Spectral type & \multicolumn{1}{l|}{$v_{rad}$ (km~s$^{-1}$)} & \multicolumn{1}{l|}{error (km~s$^{-1}$)} & \multicolumn{1}{l|}{$p_\mathrm{Member}$} & Distance (kpc) \\ 
		\hline\\ [-8pt]
		HD 326331   & O8 IVn((f)) & $-32.7^{[1]}$                              & 9                                 & 92\%  & $1.41_{-0.04}^{+0.04}$             \\
		HD 152248   & O7 Iabf + O7 Ib(f) & $-42.3^{[2]}$                              & 1.3                               & 73\%    &$1.45_{-0.06}^{+0.07}$           \\
		HD 152219   & O9.5 III(n) & $-34.3^{[3]}$                              & 1.5                               & 91\%  & $1.78_{-0.17}^{+0.21}$            \\
		HD 152270   & WC7 + O5-8 & $-44.0^{[4]}$                              & 2                                 & 69\%  & $1.50_{-0.06}^{+0.06}$            \\
		HD 326329   & O9.7 V & $-26.4^{[1]}$                              & 2.7                               & 83\% & $1.61_{-0.05}^{+0.05}$             \\
		CPD-41 7721 & O9.5 V & $-25.5^{[1]}$                              & 9.8                               & 62\%  & $1.64_{-0.06}^{+0.06}$            \\
		HD 152314   & O9 IV & $-28.5^{[1]}$                              & 7.8                               & 75\% & $1.39_{-0.05}^{+0.05}$             \\ \hline \\ [-8pt]
		HD 153919   & O6.5 Iaf+ & $-64.5^{[5]}$                                & 1.5                         &   & $1.58_{-0.06}^{+0.07}$     
	\end{tabular}
	\label{tab:radvel}
\end{table*}

\subsubsection{Radial velocity of HD 153919}

The determination of the radial velocity of HD153919 is complicated as this Of supergiant has a dense, supersonically expanding stellar wind and is in a binary system. \citet{Hutchings1974} measured the velocity of multiple spectral lines and found it to be $-60$~km~s$^{-1}$. We adopt the value of $-64.5 \pm 1.5$~km~s$^{-1}$ from Gies (1987), see also Kaper et al. (1994). The radial-velocity amplitude due to the binary motion is $20.6 \pm 1.0$~km~s$^{-1}$ \citep{Hammerschlag2003}.

\subsubsection{The space velocity of HD 153919}

The relative velocity of HD 153919 with respect to the parent cluster NGC 6231 thus becomes $v_{\rm sys} = 63 \pm 5$~km~s$^{-1}$. In the supernova scenario the runaway velocity provides a constraint on the amount of mass lost from the system during the supernova.

\subsection{The kinematical age of 4U 1700-37}

\begin{figure*}[h]
	\centering
	\includegraphics[width=\linewidth]{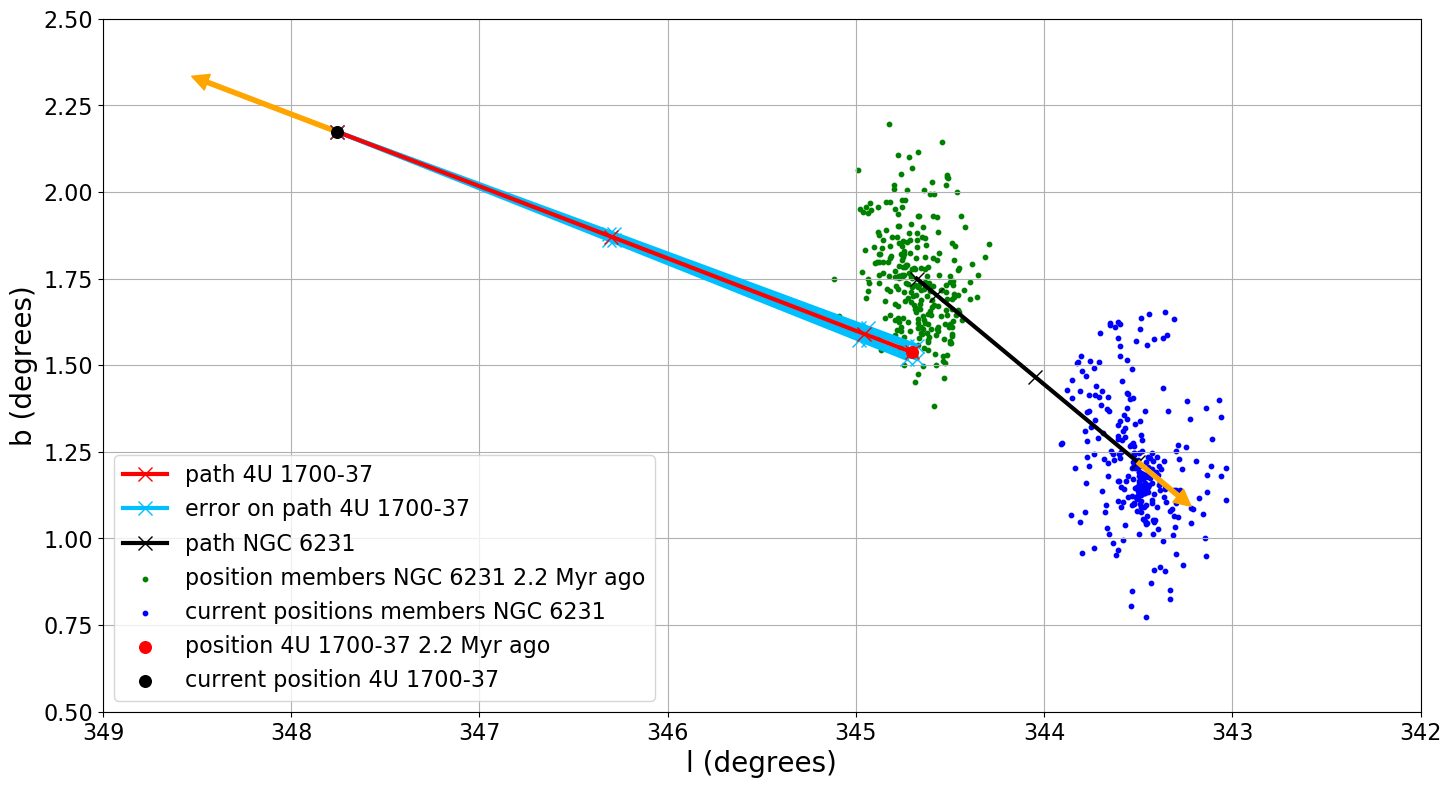}
	\caption{Location of HD 153919 / 4U 1700-37 and the individual members of NGC 6231 now (blue) and 2.2~Myr ago (green). The orange arrows indicate the movement of HD 153919 and NGC 6231 in 0.5~Myr, respectively.}
	\label{fakepath}
\end{figure*}

\begin{figure*}[h]
	\centering
	\includegraphics[width=\linewidth]{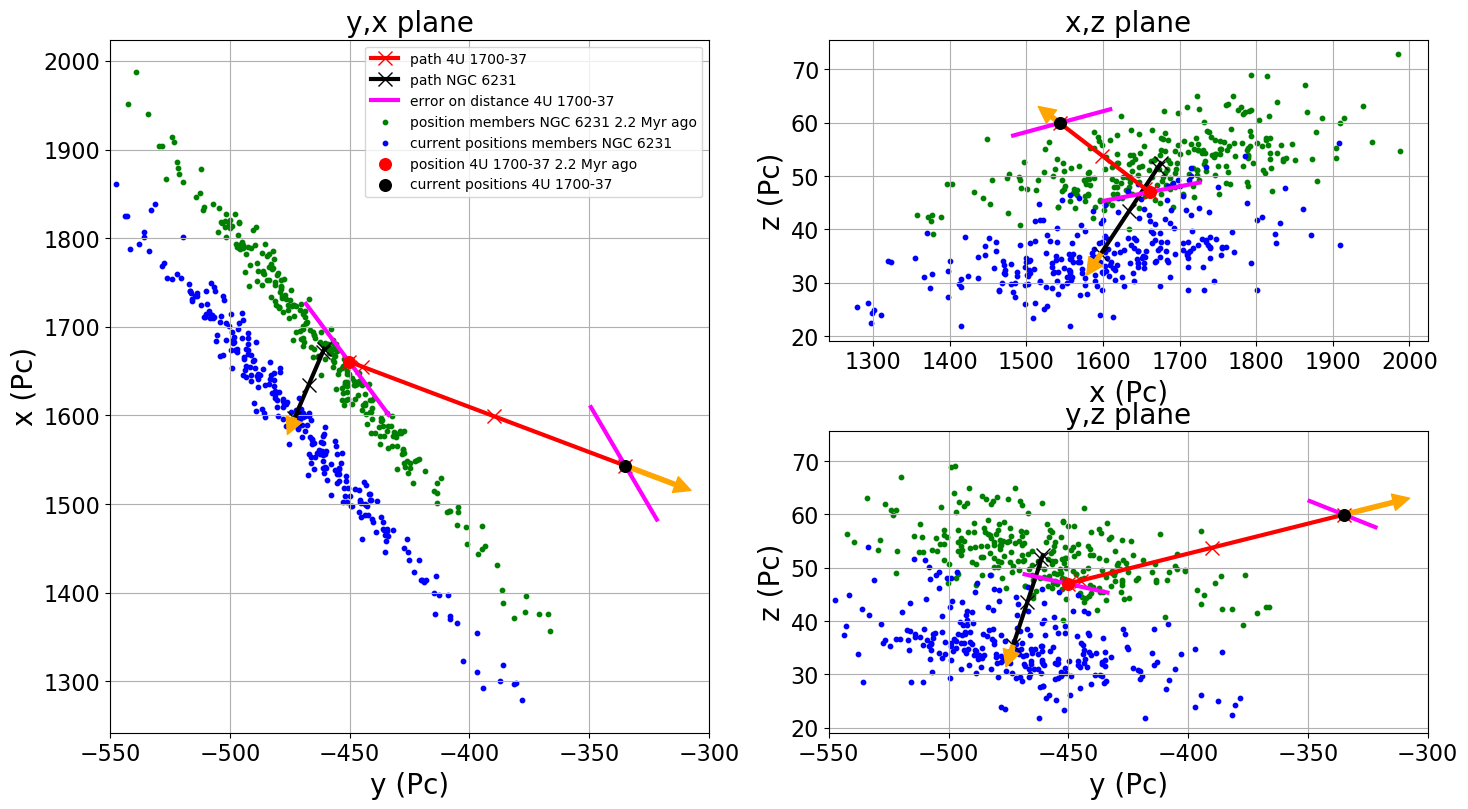}
	\caption{The position of HD 153919 and that of the individual members of NGC 6231 now (blue) and 2.2 Myr ago (green) in Galactic Cartesian coordinates. The uncertainty in the distance of HD153919 is represented by the magenta lines. The orange arrows show the movement of HD 153919 and NGC 6231 in 0.5 Myr.}
	\label{3Dpath}
\end{figure*}

The path of HD 153919/4U 1700-37 relative to NGC 6231 has been reconstructed to determine the location of the closest encounter and the time that has passed since then: the kinematical age. The coordinates, proper motions, distances, the average radial velocity of 7 likely members of NGC 6231, and the radial velocity of HD 153919/4U 1700-37 (Tab. \ref{tab:radvel}) were used to project the path on different sky planes: in Galactic coordinates $l$ and $b$ (Fig.~\ref{fakepath}) and in Cartesian coordinates $x$, $y$, and $z$ (Fig.~\ref{3Dpath}). 

It becomes clear from Fig.~\ref{fakepath} that, projected on the sky, HD 153919/4U 1700-37 was positioned in the outskirts of NGC 6231 $2.2 \pm 0.1$~Myr ago.  Fig.~\ref{3Dpath} shows that, in the radial direction, the uncertainty in the distance of HD153919 (magenta error bar) is large. This is also reflected by the elongated shape of the cluster in the radial direction. 

\section{Discussion}
\label{chap:HMXB|discussion}
From the previous sections we conclude that 2.2~Myr ago the HMXB HD 153919 / 4U 1700-37 escaped with a space velocity of 63~km~s$^{-1}$ from the outskirts of NGC 6231, the 4.7~Myr young cluster in the center of the OB association Sco~OB1. In this section we discuss the implications of the reported astrometric solution, propose scenarios to explain the evolutionary history of the system, and consider the physical nature of the compact object.

\subsection{The astrometric solution}

\textit{Gaia} determines accurate parallaxes and proper motions, but the errors on the astrometric parameters increase with distance, and the distance to HD 153919/4U 1700-37 and the cluster NGC 6231 is relatively large (1.6~kpc). Fig.~\ref{3Dpath} shows that many members of NGC 6231 appear to be up to 0.3 kpc away from the center of NGC 6231 in the radial direction. The elongated shape that the cluster seems to have is a consequence of the uncertainty in the distance determination for the individual members. If we assume the cluster to be spherical, we arrive at a physical diameter of about $20$~pc. 

In Fig.~\ref{fakepath} the cluster seems to be more compact 2.2~Myr ago than today. To investigate whether this is the result of a physical rather than an apparent expansion, we calculated the average angular distance of the members to the center of the cluster in degrees during the past 5~Myr in steps of 0.1 Myr and corrected for the projection effect by the varying distance (the cluster is approaching us). The minimum average angular distance of the cluster members to the center is reached 1.8~Myr ago (0.19$\degree$), but the difference to the current average distance (0.21$\degree$) is small. Therefore, we cannot conclude that NGC 6231 has significantly expanded since its formation about 4.7~Myr ago.

One may naively expect that HD 153919 / 4U 1700-37 would originate from the center of NGC 6231 since the progenitors of 4U 1700-37 and HD 153919 were likely both (very) massive stars (cf.\ Sect.~\ref{sec:prog}). The reconstruction of its path indicates that the binary was ejected from the outskirts of the cluster, which would not necessarily be the region with the highest stellar density for which the dynamical ejection scenario is the most efficient \citep[e.g.][]{Fujii2011, Oh2016}.

We conclude that HD 153919 / 4U 1700-37 is a runaway system that can be traced back to NGC 6231, its parent cluster. The current membership and age of NGC 6231, still including a number of massive stars, allows some stars to have been massive enough to have already ended their evolution \citep{Sung2013, Feinstein2003}. There is evidence for the occurrence of a supernova explosion in NGC 6231 \citep{Feinstein2003}, but it is unclear whether this event has a relation with the formation of 4U~1700-37. 

\subsection{Possible progenitor systems}
\label{sec:prog}

We can use the parameters of HD 153919 / 4U 1700-37 to constrain some of the characteristics of the progenitor system. \citet{Nelemans1999} derived a relation between the mass lost during a supernova explosion ($\Delta M$) and the space velocity $v_{\rm sys}$ (assuming only a Blaauw kick):

\begin{equation}
	\bigg(\frac{\Delta M}{{M}_\odot}\bigg) = \bigg(\frac{v_{\rm sys}}{213 \, {\rm km/s}}\bigg)\bigg(\frac{ M}{{M}_\odot}\bigg)^{-1}\bigg(\frac{P_{\rm cir}}{\rm day}\bigg)^{1/3}\bigg(\frac{M+m}{{M}_\odot}\bigg)^{5/3} \, ,
	\label{eq:nele}
\end{equation}

\noindent where $M$ is the present mass of HD 153919: 60 \(\textup{M}_\odot\), $m$ is the mass of the compact object in 4U 1700-37: 2.54 \(\textup{M}_\odot\), and $P_{\rm cir}$ is the orbital period after recircularization of the orbit. We neglected the eccentricity to be able to use this equation with $P_{\rm cir}$ of 3.41 days. With $v_{\rm sys}$ of 63~km~s$^{-1}$, this yields a $\Delta$M of 7.4 \(\textup{M}_\odot\). This means that the core of the progenitor of 4U1700-37 ($M_1^{\rm pre-CC}$) was about 7.5~\(\textup{M}_\odot\) after the RLOF phase and before the supernova. Such a core would be too small for a primary star with M~$> 60$~M$_{\odot}$ \citep[e.g.][]{Brott2011}, suggesting that the companion star HD153919 has accreted mass before the end of the main sequence of the progenitor of the compact object \citep[e.g.][]{Belczynski2008}, that is case A mass transfer \citep{Kippenhahn1967}. 

\citet{Renzo2019} have described an analytical method to probe the change in orbital separation using the masses of the primary and secondary before and after Roche-lobe overflow. This was originally designed to treat case~B RLOF, i.e. mass transfer after a He core is well established in the donor. While this might not be the case for 4U 1700-37, we proceed to demonstrate that assuming this scenario leads to inconsistent results.

We start with the assumption that the primary loses its complete envelope after mass transfer:

\begin{equation}
	M_1^{\rm pre-CC} \simeq M_1^{\rm ZAMS} - \mu_{\rm env}M_1^{\rm ZAMS} \, ,
	\label{eq:prima}
\end{equation}

\noindent where  $M_1^{\rm ZAMS}$ is the initial mass of the primary and $\mu_{\rm env}$ is the fraction of the primary that is the envelope.
We rewrite to get:

\begin{equation}
	\mu_{\rm env} \simeq \frac{M_1^{\rm ZAMS} - M_1^{\rm pre-CC}}{M_1^{\rm ZAMS}} \, .
\end{equation}

\noindent Then for the secondary we have:

\begin{equation}
	M_2^{\rm pre-CC} \simeq M_2^{\rm ZAMS} + \beta_{\rm RLOF}\mu_{\rm env}M_1^{\rm ZAMS} \, ,
\end{equation}

\noindent where $\beta_{\rm RLOF}$ is the fraction of the envelope of the primary that is accreted by the secondary.
We can rewrite this to:

\begin{equation}
	\beta_{\rm RLOF} \simeq \frac{M_2^{\rm pre-CC} - M_2^{\rm ZAMS}}{\mu_{\rm env}M_1^{\rm ZAMS}} \, .
\end{equation}

\noindent Since we know that $M_1^{\rm pre-CC} \simeq$ 10 ${M}_\odot$ and $M_2^{\rm pre-CC} \simeq$ 60 ${M}_\odot$, filling in $M_1^{\rm ZAMS}$ and $M_2^{\rm ZAMS}$ gives us values of $\mu_{\rm env}$ and $\beta_{\rm RLOF}$. If we find values that are physically comfortable we can proceed to calculate the orbital separation of the progenitor system. Assuming a symmetrical supernova that does not change the orbital separation we assume $a_{\rm pre-CC}$ to be the current orbital separation of 38 ${R}_\odot$. For the orbital separation of the progenitor system ($a_{\rm ZAMS}$) we can write:

\begin{equation}
	a_{\rm ZAMS} = a_{\rm pre-CC}\bigg(\frac{M_1^{\rm pre-CC}M_2^{\rm pre-CC}}{M_1^{\rm ZAMS}M_2^{\rm ZAMS}}\bigg)^2 \bigg(\frac{M_1^{\rm ZAMS}+M_2^{\rm ZAMS}}{M_1^{\rm pre-CC} + M_2^{\rm pre-CC}} \bigg)^{2 \gamma_{\rm RLOF}+1} \, ,
\end{equation}

\noindent where $\gamma_{\rm RLOF}$ is the specific angular momentum of the matter that leaves the system. For simplicity, we assume $\gamma_{\rm RLOF}=1$, i.e. mass not accreted during the RLOF takes away the specific orbital angular momentum of the binary \citep{Dominik2012}. In reality, this parameter could also vary during the mass transfer phase.

\begin{table}[h]
	\centering
	\caption{Scenarios that resulted in the largest orbital separation for $\gamma_{\rm RLOF}$ = 1, 2, 3 for this system as described in Sect.\ref{analyt}. Given are the current masses of the members of the binary: $M_2^{\rm pre-CC}$ = 60 ${M}_\odot$ and $M_1^{\rm pre-CC}$ = 2.54 + 7.4 $\simeq$ 10 ${M}_\odot$, their relative velocity to NGC 6231: $v_{\rm sys}$ = 63~km~s$^{-1}$, their current orbital period: $P_{\rm cir}$ = 3.41 days and orbital separation: $a_{\rm pre-CC}$ = 38 ${R}_\odot$. The initial mass $M_1^{\rm ZAMS}$ ranges from 30 to 80 ${M}_\odot$ and $M_2^{\rm ZAMS}$ ranges from 10 to 50 ${M}_\odot$. Only solutions with a high $\gamma_{\rm RLOF}$ and a low $\beta_{\rm RLOF}$, so when a lot of angular momentum leaves the system due to nonconservative mass transfer, seem to give a $a_{\rm ZAMS}$ that could be large enough to avoid merging of the system. For comparison: HD 153919 with a mass of 60 ${M}_\odot$ has a radius of $25 {R}_\odot$.}
	
	\begin{tabular}{l|lll}
		$\gamma_{\rm RLOF}$ & \multicolumn{1}{l|}{1} & \multicolumn{1}{l|}{2} & 3    \\ \hline
		$M_1^{\rm ZAMS}$ (${M}_\odot$)    & 56                     & 80                     & 80   \\
		$M_2^{\rm ZAMS}$ (${M}_\odot$)   & 14                     & 20                     & 50   \\
		$M_1^{\rm env}$ (${M}_\odot$) & 46                     & 70                     & 70   \\
		$\mu_{\rm env}$    & 0.82                   & 0.88                   & 0.88 \\
		$\beta_{\rm RLOF}$  & 1                      & 0.57                   & 0.14 \\
		$a_{\rm ZAMS}$ (${R}_\odot$)    & 22                     & 31                     & 64  
	\end{tabular}
	
	\label{test}
\end{table}

To find out whether we can produce a system like HD 153919 / 4U 1700-37, we tested values of $M_1^{\rm ZAMS}$ ranging from 30 to 80 ${M}_\odot$ and $M_2^{\rm ZAMS}$ ranging from 10 to 50 ${M}_\odot$, with a $\gamma_{\rm RLOF}$ of 1, 2 and 3. A $\gamma_{\rm RLOF}$ of 2 or 3 seems unlikely, since this would mean that the system suffers extreme angular momentum losses. The results of these tests are displayed in Tab.~\ref{test}. Only solutions with a high $\gamma_{\rm RLOF}$ and a low $\beta_{\rm RLOF}$, so when a lot of angular momentum leaves the system due to nonconservative mass transfer, seem to give a $a_{\rm ZAMS}$ that could be large enough to avoid merging of the system. For comparison: HD 153919 with a mass of 60 ${M}_\odot$ has a radius of 22~${R}_\odot$. This result is at least in part due to our assumption of taking the present-day separation as pre-CC separation and assume that the explosion does not change it.

\label{analyt}

\subsubsection{Binary evolution channels}

With the orbital parameters of HD 153919/4U 1700-37, the current mass of HD 153919 and 4U 1700-37, the space velocity of HD 153919 / 4U1700-37 relative to NGC 6231 and the determined lower limit of the mass of the progenitor of 4U 1700-37, population synthesis can be used to constrain the parameters of the progenitor system. The analytical approach in Sect.~\ref{analyt} showed that constraining the parameters of the progenitor system is not an easy task, as also pointed out before by \citet{Ankay2001}.

If we assume conservative mass transfer, the effect of the mass transfer on the orbital separation can be approximated by:

\begin{equation}
	\frac{\dot{a}}{a} = 2\bigg(\frac{M_d}{M_a} - 1\bigg)\frac{\dot{M_d}}{M_d} \, .
	\label{eq:mass}
\end{equation}

This tells us that because $\dot{M_d} < 0$ the orbit shrinks ($\dot{a} < 0$) as long as $M_d > M_a$ but when the mass ratio flips and $M_d < M_a$, the orbit will expand ($\dot{a} > 0$) a lot more than it shrank before the mass ratio flipped.

\paragraph{Case A mass transfer}

If we assume case A mass transfer, which overall leads to more conservative mass transfer because the mass transfer rate is relatively low and proceeds on a nuclear timescale, the orbit widens a lot as described above and the binary ends with a large orbital separation which does not fit our scenario well (assuming $a_{\rm pre-CC} = a_{\rm today}$).

\paragraph{Case B mass transfer}

If we assume case B mass transfer, which happens when the donor has a well-established He core and fills its Roche lobe because it expands significantly, the mass transfer rate is way higher than with case A mass transfer. This generally means that mass transfer is less conservative. This leads to a more compact binary, but now a different problem emerges. A supergiant, by definition, depleted the hydrogen in its core and thus has a heavy helium core after the mass transfer phase, which would probably produce a black hole after the core-collapse that is way more massive than what we have observe for 4U 1700-37 (2.54 \(\textup{M}_\odot\)). 

\paragraph{A possible scenario}

With some fine tuning one can create systems similar to what we observe. The most probable scenario includes case A mass transfer. An asymmetric supernova explosion with a very specific kick direction and amplitude can solve the complication of the widening of the binary \citep{Wongwathanarat2013, Janka2013, Janka2017}. If the kick is directed outward and kicks the compact object toward its companion, then the orbital separation can shrink enough to reproduce the observed system. 

\subsubsection{Asymmetric supernova kicks}

The discovery of radio pulsars with an extremely high space velocity exceeding in some cases 1000~km~s$^{-1}$ \citep[e.g][]{Cordes1993} is best explained with an asymmetric supernova kick. The mechanism behind supernova kicks is not fully understood. If a supernova fails, most or all matter will fall back onto the core of the star and a black hole forms. It is thought that this happens if a very massive single star undergoes a supernova explosion and the kick received here is thought to be non to little since no matter is ejected and no extra velocity is needed to conserve momentum. Black-hole systems like Cyg X-1 \citep{Mirabel2003} have low space velocities, which can be understood in terms of a failed supernova leading to fall back and the formation of a black hole, and no mass loss from the system. However, the low-mass black-hole system 4U1957+11 may have a high space velocity \text{\citep{Maccarone2020}}.

In order for a supernova explosion not to fail, it is thought that asymmetry plays a major role. The stellar plasma does not propagate outwards as fast as with a symmetric outflow because it is not accelerated only in the radial direction. This causes the density of the material to be high enough to absorb a big part of the energy carried away by the neutrino flow and accelerate outwards.

\citet{Janka2013} explains the supernova kick with a so-called tug-boat mechanism in which the most massive part of the supernova ejecta moves out slower than the rest of the ejecta because of inertia, and pulls the neutron star gravitationally giving it a velocity in that direction. The faster outflow, that causes most of the explosive nucleosynthesis, would then be situated at the other side of the core. Whether (low mass) BHs can receive kicks remains an open question \citep{Repetto2012, Mandel2016, Janka2017, Atri2019}.

\subsection{The nature of 4U 1700-37}

It is not clear whether 4U 1700-37 is a neutron star or a black hole. It emits in X-rays but does not show significant X-ray pulsations identifying it as a neutron star. \citet{White1983} states that it is possible that the neutron star undergoes spherical accretion instead of accretion along the magnetic field lines. This could be related to the accretion mechanism. Wind accretion, rather than Roche-lobe overflow, may not lead to the formation of an accretion disk and thus give rise to spherical accretion such that X-ray pulsations are not observed. Alternatively, the spin and magnetic axis of the neutron star could be aligned.

The mass of 4U 1700-37 that is estimated by \citet{Hammerschlag2003} of 2.54 $\pm$ 0.27 ${M}_\odot$ is at the high end of the mass range observed for neutron stars (2 ${M}_\odot$), but lower than than that of the lowest mass black holes (4--5 ${M}_\odot$) observed in an X-ray binary system \text{\citep[e.g.,][]{Orosz2003}}. The X-ray spectrum of 4U 1700-37 shows characteristics of a neutron star \citep{Seifina2016, White1983}.
Interestingly, LIGO/Virgo have recently announced the discovery of GW190412 \citep{Abbott2020} which involved a compact object of similar mass and unknown nature. While the metallicity of 4U1700-37 is probably different from the metallicity of the progenitor of GW190412, we suggest that 4U1700-37 is a possible analog of its progenitor. Thus studying this system might shed light on the evolution of GW progenitors. 

Perhaps the strongest argument in favor of a neutron star is that the observed runaway velocity indicates that a lot of material is lost from the system during the supernova explosion. However, it is unclear whether a low mass BH might receive a similar kick amplitude at birth.

The high space velocity of the system and the dense stellar wind are expected to result in the formation of a wind bow shock such as observed in Vela~X-1 \citep{Kaper1997} and 4U1907+09 \cite{Gvaramadze2011}. Such a wind bow shock is not observed for 4U1700-37. However, only a fraction of the OB runaway stars produces a wind bow shock \citep{Huthoff2002}. This may be either due to the low density or high temperature of the ambient medium, or both.

\begin{acknowledgements}
    We acknowledge an anonymous referee who provided constructive remarks that helped to improve the quality of the paper. We want to thank Jari van Opijnen, Mark Snelders, Rob Farmer and Selma de Mink for interesting discussions. This research has been supported by NOVA.
    
    This work has made use of data from the European Space Agency (ESA) mission
    {\it Gaia} (\url{https://www.cosmos.esa.int/gaia}), processed by the {\it Gaia}
    Data Processing and Analysis Consortium (DPAC,
    \url{https://www.cosmos.esa.int/web/gaia/dpac/consortium}). Funding for the DPAC
    has been provided by national institutions, in particular the institutions
    participating in the {\it Gaia} Multilateral Agreement.
    
\end{acknowledgements}

%
\bibliographystyle{aa} 
\bibliography{hmxb} 
%
\begin{appendix} 
\section{ADQL selection candidate members NGC 6231}
\label{ADQL}

\begin{verbatim}
	SELECT designation, source_id, 
	ra, ra_error, dec, dec_error, parallax, 
	parallax_error, parallax_over_error, 
	pmra, pmra_error, pmdec, pmdec_error, 
	ra_dec_corr, ra_parallax_corr, ra_pmra_corr, 
	ra_pmdec_corr, dec_parallax_corr, dec_pmra_corr, 
	dec_pmdec_corr, parallax_pmra_corr, 
	parallax_pmdec_corr, pmra_pmdec_corr, 
	r_est, r_lo, r_hi, 
	astrometric_primary_flag, duplicated_source, 
	radial_velocity, radial_velocity_error, 
	phot_g_mean_flux, phot_g_mean_flux_error, 
	phot_g_mean_mag, 
	phot_bp_mean_flux, phot_bp_mean_flux_error, 
	phot_bp_mean_mag, 
	phot_rp_mean_flux, phot_rp_mean_flux_error, 
	phot_rp_mean_mag, 
	phot_bp_rp_excess_factor, bp_rp, bp_g, g_rp, 
	phot_variable_flag, 
	l, b, teff_val, 
	teff_percentile_lower, teff_percentile_upper, 
	a_g_val, a_g_percentile_lower, 
	a_g_percentile_upper, 
	e_bp_min_rp_val, 
	e_bp_min_rp_percentile_lower, 
	e_bp_min_rp_percentile_upper, 
	radius_val, radius_percentile_lower, 
	radius_percentile_upper, 
	lum_val, lum_percentile_lower, 
	lum_percentile_upper, 
	astrometric_chi2_al, astrometric_n_good_obs_al
	
	FROM external.gaiadr2_geometric_distance 
	JOIN gaiadr2.gaia_source USING (source_id) 
	WHERE l >= 342.5 AND l <= 344 
	AND b >= 0 AND b <= 2 AND 
	parallax >= 0.4 AND parallax <= 1
\end{verbatim}

\section{Membership probability thresholds for NGC 6231}
\label{appendix:pmin}
\begin{table}[h!]
	\centering
	\caption{The number of candidates $n_{*}$ in relation with the probability threshold $p_{min}$ for NGC 6231.}
	\begin{tabular}{c|c}
		\hline
		$p_{min}$ & $n_{*}$ \\
		\hline
		0.50 & 786 \\
		0.55 & 763 \\
		0.60 & 731 \\
		0.65 & 700 \\
		0.70 & 667 \\
		0.75 & 629 \\
		0.80 & 546 \\
		0.85 & 430 \\
		0.90 & 268 \\
		0.95 & 0   \\
		\hline
	\end{tabular}
	\label{tab:pmin_n_relation}
\end{table}

\section{Isochrone parameters}

\subsection{PARSEC isochrone parameters}
\label{PARSEC}

Thermal pulse cycles included\\ 
On RGB, assumed Reimers mass loss with efficiency eta=0.2\\
Photometric system: Gaia's DR2 G, G\_BP and G\_RP (Vegamags, Gaia passbands from Maiz-Apellaniz and Weiler 2018)\\
Using YBC version of bolometric corrections as in Chen et al. (2019)\\
O-rich circumstellar dust ignored\\
C-rich circumstellar dust ignored\\
IMF: chabrier\_lognormal\_salpeter \\
Kind of output: isochrone tables\\

\subsection{Mist isochrone parameters}
\label{MIST}

MIST version number  = 1.2\\     
MESA revision number =     7503\\
photometric system   = UBV(RI)c, 2MASS, Kepler, Hipparcos, Gaia (Vega)\\                 
Yinit = 0.2703     Zinit = 1.42000E-02 [Fe/H] = 0  [a/Fe] = 0  v/vcrit = 0\\


\end{appendix}

\end{document}